\documentclass[12pt]{article}
\usepackage{amsmath,amssymb}
\usepackage{bm}
\usepackage{color}
\usepackage{cite}
\usepackage{mathtools}

\usepackage{here}
\usepackage {cancel}

\usepackage{multirow}

\usepackage{dsfont}

\usepackage{here}

\begin{document}
	
	\title{
		\begin{flushright}
			\ \\*[-80pt]
			\begin{minipage}{0.2\linewidth}
				\normalsize
			\end{minipage}
		\end{flushright}
	{\Large \bf
		$A_4$ modular invariance and the strong CP problem
		\\*[20pt]}}
	\author{
		~S. T.~Petcov $^{1,2}$\footnote{Also at:
			Institute of Nuclear Research and Nuclear Energy,
			Bulgarian Academy of Sciences, 1784 Sofia, Bulgaria.}~  
		and~ M. Tanimoto $^{3}$
		\\*[20pt]
		{
			\begin{minipage}{\linewidth}
				$^1${\it \normalsize
					INFN/SISSA, Via Bonomea 265, 34136 Trieste, Italy} \\*[5pt]
				$^2${\it \normalsize Kavli IPMU (WPI), UTIAS, The University of Tokyo, 
					Kashiwa, Chiba 277-8583, Japan}
				\\*[5pt]
				$^3${\it \normalsize
					Department of Physics, Niigata University, Ikarashi 2, 
					Niigata 950-2181, Japan} 
				\\*[5pt]
			\end{minipage}
		}
		\\*[40pt]}
	
	\date{
		\centerline{\small \bf Abstract}
		\begin{minipage}{0.9\linewidth}
			\medskip
			\medskip
			\small
We present simple effective 
theory of quark masses, mixing and CP violation 
with level $N=3$ ($A_4$) modular symmetry, which 
provides solution to the 
strong CP problem without the need for an axion. 
The vanishing of the strong CP-violating phase $\bar \theta$	
is ensured by assuming CP to be a fundamental symmetry of the 
Lagrangian of the theory. The CP symmetry is broken spontaneously 
by the vacuum expectation value (VEV) of the modulus $\tau$.  
This provides the requisite large value of the CKM CP-violating phase 
while  the strong CP phase $\bar \theta$ remains zero or is tiny.
Within the considered framework we discuss 
phenomenologically viable quark mass matrices with 
three types of texture zeros, which 
are realized by assigning both the left-handed and right-handed 
quark fields to $A_4$ singlets ${\bf 1}$, ${\bf 1'}$ and ${\bf 1''}$ 
with appropriate weights. The VEV of  $\tau$ is restricted  to 
reproduce the observed CKM parameters. We discuss
cases in which the modulus VEV is close to the fixed points 
$i$, $\omega$ and $i\infty$.  
In particular, we focus  on the VEV of $\tau$,  
	which gives the absolute minima of the supergravity-motivated
	modular- and CP-invariant potentials for the modulus $\tau$,
	so called, modulus stabilisation. We present a successful model, which is  consistent with the  modulus stabilisation  close to $\tau=\omega$.
\end{minipage}
}
	
	\begin{titlepage}
		\maketitle
		\thispagestyle{empty}
	\end{titlepage}
	

\section{Introduction}
The strong CP problem has been a puzzle in particle physics since
the QCD Lagrangian violates CP due to instanton effects
	\cite{Belavin:1975fg,Jackiw:1976pf,Callan:1976je,tHooft:1976rip}.
	The CP violation in QCD is described by the strong CP phase
\begin{align}
\bar \theta=\theta_{\rm QCD}+{\rm arg}\,{\rm det }\,[M_U M_D]\,,
\end{align}
%
where $M_U$ and $M_D$ denote the  mass matrices of the up-type 
and the down-type quarks, respectively, and $\theta_{\rm QCD}$
is the coefficient of the topological charge term
in the QCD Lagrangian:
\begin{align}
{\cal L}_{QCD}=\bar Q(i \ \cancel {D}-M_Q)Q -\frac14 {\rm Tr}\, G^2+
\theta_{\rm QCD}\frac{g_3^2}{32\pi^2}{\rm Tr}\, G\tilde G\,.
\end{align}
%
Here $G$ is the gluon field stress tensor and $Q$ is the 
multiplet of quark fields with definite masses.
While $\theta_{\rm QCD}$ and ${\rm arg}\,{\rm det }\,[M_U M_D]$
are transformed into each other via chiral transformation, the 
sum $\bar\theta$ is invariant.

 The upper-bound of  $\bar\theta$ is derived  from the experimental
upper bound on the electric dipole moment (edm) of the neutron
 as $|\bar \theta| \lesssim 10^{-10}$ \cite{Abel:2020pzs}.
 This bound is much smaller than the
  weak CP violation, for example, the Jarlskog 
rephasing invariant $\sim 10^{-5}$
  \cite{Jarlskog:1985ht,ParticleDataGroup:2022pth}.
 Therefore, the strong CP problem is the problem of understanding why 
$\bar{\theta}$  is so small.
 
 The most well known solution is the axion solution \cite{Peccei:1977hh}
 (see also, e.g., \cite{DiLuzio:2020wdo}),
where $\bar\theta$ is  a dynamical degree of freedom 
which is set to a small value by a scalar potential.
However, there have been no experimental hints for 
the existence of the axion so far.
 
Another well known solution is provided by the Nelson-Barr model
 \cite{Nelson:1983zb,Barr:1984qx},
 where the CP symmetry is violated only by the mixings 
of Standard model (SM) quarks with hypothetical extra heavy quarks,
 and the extended quark mass matrix has a special structure with vanishing 
 entries such that CP-violating terms do not contribute
 to its determinant.
Specific models, in which quark mass matrices with real determinant
 and   $\theta_{\rm QCD}=0$ still generate the weak  CP-violating (CPV) phase
 in the CKM quark mixing matrix  have been proposed in
  \cite{Dine:2015jga,Hiller:2001qg,Babu:1989rb,Bento:1991ez,Kuchimanchi:1995rp
  	,Barr:1991qx,Bonnefoy:2023afx,Antusch:2013rla,Harnik:2004su,
  	Cheung:2007bu,Vecchi:2014hpa,Berezhiani:1992pq,Valenti:2021xjp} 
as well.
 
The modular invariance opened up a new promising 
 approach to the flavour problem
of quarks and leptons \cite{Feruglio:2017spp}
(see also Refs.\cite{Kobayashi:2018vbk,Penedo:2018nmg,Novichkov:2018nkm}).
The strong CP problem has been also  discussed recently 
within the modular invariance approach in Ref.\cite{Feruglio:2023uof}.
In this study a simple effective theory of flavour and CP symmetry
breaking with  vanishing  $\bar\theta$ without the need for an axion
has been proposed.
The analysis was done 
using the level $N = 1$ full modular group SL$(2,\mathbb{Z})$.
A numerical example has shown that the modular symmetry
allows to solve the strong CP problem  and to reproduce correctly
the quark masses and  the CKM mixing matrix.
 
In this article we present alternative 
effective quark flavour models with finite modular symmetry 
of level $N=3$ ($A_4$), which provide axion-less solution 
to the strong CP problem. The $A_4$ modular symmetry 
has been extensively used for understanding 
the origins of the quark and lepton flavours
\cite{Feruglio:2017spp,Criado:2018thu,Kobayashi:2018scp,deAnda:2018ecu,Novichkov:2018yse,Okada:2018yrn,Ding:2019zxk,Okada:2019uoy,Okada:2020rjb,Okada:2020ukr,Okada:2020brs,Okada:2021qdf,Kobayashi:2021pav,Kobayashi:2022jvy,Petcov:2022fjf,Petcov:2023vws}. 
Finite groups including finite modular groups 
have been widely employed in flavour model building 
(see, e.g., \cite{Kobayashi:2018vbk,Penedo:2018nmg,Novichkov:2018ovf,Novichkov:2018nkm,Liu:2019khw,Novichkov:2020eep,Yao:2020zml,Ding:2021iqp,Novichkov:2021evw,Ding:2021zbg,Li:2021buv,deMedeirosVarzielas:2023crv}
\footnote{A rather complete list of articles on the modular 
invariance approach to lepton and quark flavour problems 
is given in \cite{deMedeirosVarzielas:2023crv}.
}
and the reviews  
\cite{Altarelli:2010gt,Ishimori:2010au,Ishimori:2012zz,Hernandez:2012ra,King:2013eh,King:2014nza,Tanimoto:2015nfa,King:2017guk,Petcov:2017ggy,Kobayashi:2022moq,Feruglio:2019ybq,Kobayashi:2023zzc}).

  If CP is a fundamental symmetry of the Lagrangian, the QCD and 
the strong CPV phases  
$\theta_{\rm QCD}$ and $\bar \theta$  will vanish.
However, in order to explain CP violation in weak interactions, 
the spontaneous breaking of the CP symmetry has to 
generate the large measured value of the CPV phase
in the CKM matrix, while at the same time 
the strong CPV phase $\bar \theta$ should still be zero 
or be tiny enough to be in agreement with experimental data on 
the edm of the neutron. In other words, we have to look for a texture with 
${\rm arg}\,{\rm det }\,[M_U M_D]=0$ with a realistic value for the 
CKM CPV phase.
Furthermore, as far as there is no  accidental cancellation between  
phases in the up-type  and down-type mass matrices,
${\rm det }\,[M_U]$ and ${\rm det }\,[M_D]$ should be real, 
and positive  by themselves.
 
 Since  constraints of $\bar\theta$ are very stringent, 
we need to take into account the  corrections to this parameter. 
The most important corrections are 
\footnote{ The corrections to $\bar \theta$ from SM
	are known to be negligible \cite{Ellis:1978hq,Khriplovich:1985jr}
} 
\cite{Antusch:2013rla,Feruglio:2023uof}:
\begin{itemize}
	\item{Higher dimensional operators 
that spoil the structure of the mass matrices.}
	\item{Corrections which are induced from supersymmetry (SUSY) breaking 
terms.}
\end{itemize}

The first point has been  addressed by introducing the modular symmetry 
within the modular invariance approach to the quark (lepton) 
flavour problem \cite{Feruglio:2017spp}, 
in which the elements of the quark (charged lepton and neutrino) mass 
matrices are modular forms of certain levels and weights that are  
holomorphic functions of the modulus $\tau$.
The modular forms are essentially  non-perturbative ones.
Once the weight of the modular form is fixed there are no the higher 
dimensional operator contributions. The symmetry is sharp.

The second point is the correction due to the SUSY breaking terms. 
This correction has been discussed in the modular symmetry
of flavours \cite{Feruglio:2023uof}.
Since the  modular symmetry   is in the
framework of  SUSY, the SUSY breaking could  contribute
to $\bar\theta$ significantly.
As far as  SUSY is unbroken, $\bar{\theta}$ cannot be generated radiatively. 
On the other hand, the SUSY breaking sector, in principle,
can  introduce new sources of CP violation which  are  model dependent.
Assuming that SUSY is broken via gauge-mediation 
or anomaly-mediation below  the mass scale of the modulus $\tau$,
the renomalization group and threshold corrections due to the SUSY breaking have same flavour and CP structure as the SM ones \cite{Hiller:2001qg}.
Therefore,  the corrections to $\bar \theta$ 
are safely under control.

In the case of $A_4$ modular symmetry 
considered by us relevant texture zeros of the quark 
mass matrices are realized if both left-handed (LH) and right-handed (RH)  
quark fields are assigned to be  $A_4$ singlets \cite{Zhang:2019ngf}.
The CP violation is generated only by the vacuum expectation value (VEV) of 
the modulus $\tau$ \cite{Novichkov:2019sqv}.
By performing statistical analysis 
we obtain the values of the VEV of $\tau$  
that allow to reproduce the observed CP violation 
in the quark sector.

It is known that the VEV of the modulus $\tau$  
could be obtained by the modulus stabilisation
analysing  relatively simple (supergravity-motivated) 
modular- and CP-invariant potentials for the modulus $\tau$. 
In the modulus stabilisation studies 
values of the VEV of $\tau$ close to 
the fixed points \cite{Novichkov:2018ovf}
$i$, $\omega$ 
and  $i\infty$ were found 
\cite{Kobayashi:2019xvz,Abe:2020vmv,Ishiguro:2020tmo,Novichkov:2022wvg,Ishiguro:2022pde,Knapp-Perez:2023nty,King:2023snq,Kobayashi:2023spx,Higaki:2024jdk}.  
In view of this we search in this work 
for examples of $A_4$ modular solutions 
to the strong CP problem for values of the VEV 
of $\tau$ close to the fixed points. 

 The paper is organised as follows.
In Section 2,  we present quark mass matrices with 
three types of texture zeros, which  satisfy ${\rm arg}\,{\rm det }\,[M_Q]=0$.
In Section 3, we propose  quark mass matrices with 
the considered texture zeros
in the case of $A_4$ modular symmetry and 
discuss their CP violating phase structure. 
In Section 4, we show numerical examples which allow 
to reproduce the  observed  quark masses, CKM mixing angles and 
 CP-violating phase for the VEV of  $\tau$ close to the fixed points
$i$, $\omega$ and  $ i\infty$.
In Section 5,
	we focus  on the VEV of $\tau$,  
	which gives the absolute minima of the supergravity-motivated
	modular- and CP-invariant potentials for the modulus $\tau$,
	so called, modulus stabilisation. We present a successful model, which is  consistent with the  modulus stabilisation  close to $\tau=\omega$	\cite{Novichkov:2022wvg}.
Section 6 is devoted to the summary and discussions.
In Appendix \ref{modular-forms} the $A_4$
modular forms of weights up to 16 are presented. 
In Appendix \ref{Appen-inputs} we list the observed values of the 
quark masses, the CKM elements
and CPV phase, which serve as input for our numerical analyses.
In Appendix \ref{Appen-texture}  
we present numerically three examples of phenomenologically viable 
quark mass matrices with texture zeros that allow 
to describe the data on quark masses, mixing and CP violation.


%
\section{Texture zeros and Real $\bf {\bf det }\,[M_Q]$}
\label{sec:TZ}
%
%

The texture zeros approach has a long history.
In Ref.\cite{Weinberg:1977hb} Weinberg, 
assuming the existence of two families of quarks,
considered a mass matrix for the down-type quark 
sector with zero  (1,1) entry in the basis in which the 
up-type quark mass matrix is diagonal.  
He further supposed that the down-type quark mass matrix is
symmetric. In this case the number 
of free parameters is reduced to only two 
and hence he succeeded to predict the Cabibbo angle  to be  $\sqrt{m_d/m_s}$,
which is the so-called Gatto, Sartori, Tonin relation \cite{Gatto:1968ss}. 
Fritzsch extended the above approach to the three
family case \cite{Fritzsch:1977vd,Fritzsch:1979zq}. 
Ramond, Roberts and Ross presented 
a systematic analysis with  four or five zeros
for symmetric or hermitian quark mass matrices \cite{Ramond:1993kv}.
Their textures are not viable today since they cannot describe 
the current rather precise data on the CKM quark mixing 
matrix. However, the texture zero approach to the quark mass matrices 
is still promising \cite{Fritzsch:2002ga,Xing:2015sva}
\footnote{By making a suitable weak basis transformation, one can obtain 
some  sets of zeros of the quark mass matrices. This issue was discussed  
in Refs. \cite{Branco:1988iq,Branco:1994jx,Branco:1999nb}}.

A systematic study of texture zeros has been presented  for the down-type  
quark mass matrix in the basis of diagonal up-type quark mass matrix 
in Ref.\cite{Tanimoto:2016rqy} from  
the standpoint of ``Occam's Razor approach'' \cite{Harigaya:2012bw},
 in which a minimum number of parameters is  allowed.
The down-type quark mass matrix was arranged to have the minimum 
number of parameters by 
  setting three of its elements to zero, while at the 
same time requiring that it
describes successfully the CKM mixing and CP violation without 
assuming it to be symmetric or hermitian. 
Some of the texture zeros considered in  Ref. \cite{Tanimoto:2016rqy} 
lead to real ${\rm det }\,[M_Q]$.
For the purpose of our study, where the modular invariance
of the quark mass matrices determines 
the texture zeros,  
we discuss  three sets of texture zeros for 
both the down-type and up-type 
mass matrices, $M_Q$, which are texture zeros of the 
corresponding quark Yukawa couplings $Y_Q$ 
\footnote{There 
are thirteen available textures with three zeros,
of which six textures lead to real 
${\rm det }\,[M_Q]$ \cite{Tanimoto:2016rqy}.
They give three different relations 
between the CKM angles and the quark masses
as seen in Table 1 of Ref.\cite{Tanimoto:2016rqy}.
The three textures in Eq. \eqref{textures}  are 
representatives of the indicated six textures.
}: 
$M_{Q}=v_Q Y_{Q}$, $Q=D,U$, $v_{D,U}$ being the VEVs of the down-type and up-type doublet Higgs fields.
In the right-left (RL) convention for the mass 
matrices $(M_{Q}) _{RL}=v_Q (Y_{Q})_{RL}$, 
which we are going employ in our analysis,
the three sets of textures we were referring to above are:
\begin{align}
 &(1): \quad Y_Q = v^{-1}_Q\,M_{Q} = 
\begin{pmatrix}
0  & 0 & a_Q  \\
0&b_Q & c_Q \\
d_Q& e_Q& f_Q  \, e^{i \varphi_Q}\end{pmatrix}_{RL},
\nonumber\\
&(2): \quad Y_Q = v^{-1}_Q\,M_{Q} 
=\begin{pmatrix}
0  & 0 & a_Q  \\
b_Q& 0 & c_Q\\
e_Q& d_Q& f_Q  \, e^{i \varphi_Q}\end{pmatrix}_{RL},
\nonumber\\
&(3): \quad Y_Q = v^{-1}_Q\,M_{Q} = 
\begin{pmatrix}
a_Q &0 & 0  \\
c_Q&  b_Q &0\\
f_Q \, e^{i \varphi_Q}& e_Q& d_Q \end{pmatrix}_{RL},
\label{textures}
\end{align}
%
where $Q=D,\,U$, and $a_Q,\,b_Q,\,c_Q,\,d_Q,\,e_Q,\,f_Q$
are real coefficients.
The texture (1)  is used in Ref. \cite{Feruglio:2023uof}
for the discussion of the strong CP problem. 
The textures (2) and (3) are obtained by exchange of columns
$(1 \leftrightarrow 2)$ and $(1 \leftrightarrow 3)$ 
of the matrix (1), respectively.
The  physics (the CKM matrix) is different among them.
On the other hand, the  physics is not changed 
by the exchange of the rows of (1) since it respects only the right-handed
sector.
The CPV phases $\varphi_D$ and $\varphi_U$  are assigned 
to specific entries so that the mass matrices in Eq.\,(\ref{textures})
coincide in form with those obtained from modular invariance
in our further analysis.  As we will see, the modular invariance
constraints the down-type 
and up-type quark mass matrices we will consider  
to have each only one CPV phase 
originating from the VEV of the modulus $\tau$.
The determinants of the considered mass matrices are real:
\begin{align}
{\rm det }\,[M_Q]=\pm a_Q b_Q d_Q\,.
\label{det-texture}
\end{align}
%
We show typical examples of numerical fits for the 
three textures of Eq.\,\eqref{textures} in Appendix \ref{Appen-texture}, 
for which the input data are given in Appendix \ref{Appen-inputs}.

%
\section{Realization of texture zeros in $A_4$ modular symmetry}
\label{A4}
%
%
 We will present next modular invariant mass matrices with 
level  $N=3$  ($A_4$) modular  symmetry.
They have the texture zeros of the matrices given in 
Eq.\,\eqref{textures}.
In order to realize the desirable texture zeros, all quarks should be assigned to $A_4$ singlets.
However, the modular forms representing trivial singlets {\bf 1}
are just equivalent to the  Eisenstein series,
which correspond to the modular forms of level $N=1$ modular symmetry, 
as shown  in Eq.\,\eqref{Eisen} of Appendix \ref{modular-forms}.
An example of  modular invariant mass matrix  with level $N=1$ 
modular symmetry was presented in Ref. \cite{Feruglio:2023uof}.
In this work, we employ modular forms that are non-trivial singlets 
{$\bf 1'$} and {$\bf 1''$}  in addition to forms which are 
trivial singlets  {\bf 1}.
In order to reproduce the observed   CP violation of the quark sector,
 two singlet modular forms with same weight
{furnishing the same singlet representation}
are required 
to avoid the vanishing of 
the  CPV phase 
due to cancellation between the contributions for the 
 down-type and up-type quark mass matrices.
The smallest weight $k_Y$
at which there exists two $A_4$ modular forms of the same singlet 
representation is $k_Y=12$, but the forms are trivial {\bf 1} singlets. 
Two non-trivial  {$\bf 1'$} singlet modular forms 
of the same weight exist at  $k_Y=16$ and at larger weights,
as seen in  Appendix \ref{modular-forms}.
 In this section, using two ${\bf 1'}$ singlet modular forms with 
weight $k=16$ together with additional singlet modular forms, 
we construct 
quark mass matrices with $A_4$ modular symmetry, 
which have the general forms  given in Eq.\,(\ref{textures}).  
   
%
%
\subsection{Quark mass matrix of model (1) and its CPV phase structure}
In order to keep  ${\rm det }\,[M_D M_U]$ to be real,
 the following condition of the weights are required  \cite{Feruglio:2023uof}:
  \begin{eqnarray}
\sum^3_{i=1}  (2\,k_{Qi} + k_{d_i}^c + k_{u_i}^c) = 0\,,
 \label{weight-condition}
 \end{eqnarray}
where $k_{Qi}$ and  $k_{d_i}^c (k_{u_i}^c)$
denote weights for the left-handed quarks and right-handed
 down (up)-type quarks, respectively.
 The weights of Higgs are set to be zero.
 The condition of Eq.\eqref{weight-condition}  guarantees that the modular symmetry has
  no QCD anomaly.

We  assign the $A_4$ representations and the weights 
for  quarks and Higgs as follows:
\begin{itemize}
  \item{quark doublet (left-handed) $Q_1 =(d,u)_L,Q_2=(s,c)_L,Q_3=(b,t)_L$: $A_4$ singlets
  	$(1,\,1,\,1'')$ with weight $(-8,-4,8)$.}
  \item{quark singlets (righ-handed) $(d^c,s^c,b^c)$ and
  	$(u^c,c^c,t^c)$
  	: $A_4$ singlets $(1',\,1,\,1)$ with weight $(-8,4,8)$,
  	respectively.}
  \item{ Higgs fields of down-type and up-type quark sectors $H_{U,D}$: $A_4$ singlet $1$ with weight 0.}
\end{itemize}
The assigned weights satisfy the condition in Eq.\eqref{weight-condition}.
These assignments are summarized in Table \ref{tab:model-1}.
%
\begin{table}[h]
	\begin{center}
		\renewcommand{\arraystretch}{1.1}
		\begin{tabular}{|c|c|c|c|c|} \hline
			& $(d,u)_L,(s,c)_L,(b,t)_L$ & $(d^c,s^c,b^c),\,(u^c,c^c,t^c)$ &  $H_U$ & $H_D$ \\ \hline
			$SU(2)$ & 2 & 1  & 2 & 2 \\
			$A_4$ & $(1,\,1\,,1'')$ &  $(1',\,1\,,1)$ & $1$ & $1$ \\
			$k$ &$ (-8,-4,\ 8)$  &$(-8,4,\ 8)$  & 0 & 0 \\ \hline
		\end{tabular}
	\end{center}
\caption{Assignments of $A_4$ representations and weights in model (1). 
We note that the exponent in the automorphy factor in the modular 
transformations of the quark superfields we work with is $(-\,k)$.  
}
	\label{tab:model-1}
\end{table}

Taking account of the following tensor products of  $A_4$ singlets
\cite{Ishimori:2010au,Ishimori:2012zz,Kobayashi:2022moq},
\begin{align}
 {\bf 1} \otimes {\bf 1} = {\bf 1}\,, \qquad 
{\bf 1'} \otimes {\bf 1'} = {\bf 1''}\,, \qquad
{\bf 1''} \otimes {\bf 1''} = {\bf 1'}\,, \qquad
{\bf 1'} \otimes {\bf 1''} = {\bf 1}\,,
\end{align}
the superpotential terms $w_D$ and $w_U$ of the  down-type and up-type 
quark superfields with weights as given in
 Table \ref{tab:model-1} read:
\begin{align}
&w_D =
\left[a_D  d^c Q_3+ b_D  s^c Q_2+ c_D s^c Q_3 Y_{\bf 1'}^{(12)}
+d_D  b^c Q_1 +e_D b^c Q_2 Y_{\bf 1}^{(4)}
+f_D b^c Q_3\, (g_D Y_{\bf 1'{\rm A}}^{(16)}+Y_{\bf 1'{\rm B}}^{(16)}) \right] H_D\, 
,\nonumber \\
&w_U =
\left[a_U  u^c Q_3+ b_U  c^c Q_2+ c_U c^c Q_3 Y_{\bf 1'}^{(12)}
+d_U  t^c Q_1 +e_U t^c Q_2 Y_{\bf 1}^{(4)}
+f_U t^c Q_3 (g_U Y_{\bf 1'{\rm A}}^{(16)}+Y_{\bf 1'{\rm B}}^{(16)})\right] H_U\,
,\nonumber \\
\label{superpotential-1}
\end{align}
%
where $Y_{\bf r}^{(k_Y)}$
  are modular forms of ${\bf r=1,1'}$ with weight $k_Y$
as seen in Appendix \ref{modular-forms}. Parameters $a_D,\, b_D,\, c_D,\, d_D,\, e_D,\,f_D,\,g_D$ and 
$a_U,\, b_U,\, c_U,\, d_U,\, e_U,\,f_U,\,g_U$ are constants. 
These constants are real as a consequence of the imposed CP invariance 
of the Lagrangian of the theory.
We note that $w_D$ and $w_U$ involve three modular forms that are 
  {$\bf 1'$} singlets, $Y_{\bf 1'}^{(12)}$, $Y_{\bf 1'{\rm A}}^{(16)}$ 
and $Y_{\bf 1'{\rm B}}^{(16)}$, in addition to the trivial singlet one 
 $Y_{\bf 1}^{(4)}$.

The Yukawa matrices of the down-type and up-type quarks follow from the 
expressions for  $w_D$ and $w_U$ and are given by:
\begin{align}
  Y_D =
\begin{pmatrix}
0  & 0 & a_D  \\
0&b_D & c_D   Y_{\bf 1'}^{(12)}\\
d_D& e_D   Y_{\bf 1}^{(4)}&
f_D (g_D Y_{\bf 1'{\rm A}}^{(16)}+Y_{\bf 1'{\rm B}}^{(16)}) \end{pmatrix}_{RL}\,,
\quad
Y_U =
\begin{pmatrix}
0  & 0 & a_U  \\
0&b_U & c_U  Y_{\bf 1'}^{(12)}\\
d_U& e_U  Y_{\bf 1}^{(4)}&
f_U (g_U Y_{\bf 1'{\rm A}}^{(16)}+
Y_{\bf 1'{\rm B}}^{(16)}) \end{pmatrix}_{RL}\,.
\label{Ymodel-1}
\end{align}

 
 We take the modular invariant  kinetic terms simply as
 {\footnote{Possible non-minimal additions to the K\"ahler potential, 
 		compatible with the modular symmetry, may jeopardise the predictive power 
 		of the approach~\cite{Chen:2019ewa}. 
 	However, those do not affect $\arg \det M_Q$ as discussed in Ref.\cite{Feruglio:2023uof}. }
 \begin{equation}
 \sum_I\frac{|\partial_\mu\psi^{(I)}|^2}{\langle-i\tau+i\bar{\tau}\rangle^{k_I}} ~,
 \label{kinetic}
 \end{equation}
 %
 where $\psi^{(I)}$ denotes a chiral superfield with weight $k_I$,
 and $\bar\tau$ is the anti-holomorphic modulus.
 The (anti-)holomorphic modulus  is a dynamical field.
It becomes a complex number after the modulus $\tau$ takes a VEV.
 Then, one can set $\bar \tau=\tau^*$.
 
 It is important to  address  the transformation needed to get the kinetic 
 terms of matter superfields in canonical form because the
 terms  in  Eq.\,(\ref{kinetic}) are not canonical. 
The canonical form  is obtained by an overall re-normalization of 
the quark superfields of interest:
 \begin{eqnarray}
 \psi^{(I)}\rightarrow  \sqrt{(2{\rm Im}\tau_q)^{k_I}} \, \psi^{(I)}\,.
 \label{canonical}
 \end{eqnarray}
 %
 This changes some of the constant parameters in the 
quark mass matrices in the following way:
 \begin{eqnarray}
 &&c_Q \rightarrow \hat c_Q= c
_Q\, \sqrt{(2 {\rm Im} \tau)^{12} }
 =c_Q\,(2 {\rm Im} \tau)^{6},\, \nonumber\\
  &&e_Q \rightarrow \hat e_Q= e_Q\, \sqrt{(2 {\rm Im} \tau)^{4} }
 =e_Q\,(2 {\rm Im} \tau)^{2},\, \nonumber\\
  &&f_Q \rightarrow \hat f_Q= f_Q\, \sqrt{(2 {\rm Im} \tau)^{16} }
 =f_Q\,(2 {\rm Im} \tau)^{8}\,,\qquad Q=D,\,U\,.
\nonumber\\
 \label{shift}
 \end{eqnarray}
%
The constants $a_Q$, $b_Q$, $d_Q$ remain unchanged.
Thus, the mass matrices of the down-type and up-type quarks 
take the form:
\begin{align}
&  M_D =v_D
\begin{pmatrix}
0  & 0 & a_D  \\
0&b_D & c_D  (2 {\rm Im} \tau)^{6} Y_{\bf 1'}^{(12)}\\
d_D& e_D  (2 {\rm Im} \tau)^{2} Y_{\bf 1}^{(4)}&
 f_D (2 {\rm Im} \tau)^{8}(g_D Y_{\bf 1'{\rm A}}^{(16)}+
 Y_{\bf 1'{\rm B}}^{(16)}) \end{pmatrix}_{RL}\,,
\nonumber\\
&M_U =v_U
\begin{pmatrix}
0  & 0 & a_U  \\
0&b_U & c_U (2 {\rm Im} \tau)^{6} Y_{\bf 1'}^{(12)}\\
d_U& e_U (2 {\rm Im} \tau)^{2} Y_{\bf 1}^{(4)}&
f_U (2 {\rm Im} \tau)^{8} (g_U Y_{\bf 1'{\rm A}}^{(16)}+
Y_{\bf 1'{\rm B}}^{(16)}) \end{pmatrix}_{RL}\,.
\label{model-1}
\end{align}
%
The determinants of the quark mass matrices $M_D$ and $M_U$ 
given   in the preceding equation 
are real:
\begin{align}
{\rm det }\,[M_D]=-a_D b_D d_D\,,\qquad\qquad 
{\rm det }\, [M_U]=-a_U b_U d_U\,.
\end{align}
%

The phase structure of the matrices in Eq.\,\eqref{model-1} is 
rather simple. The CP violating phases originate from the 
VEV of the modulus $\tau$ via the modular forms
$Y_{\bf 1}^{(4)}$,  $Y_{\bf 1'}^{(12)}$,
$Y_{\bf 1'{\rm A}}^{(16)}$ and $Y_{\bf 1'{\rm B}}^{(16)}$.
The phases of  $Y_{\bf 1}^{(4)}$ and  $Y_{\bf 1'}^{(12)}$
in the $(2,3)$ and $(3,2)$ elements can be factorised as follows:
\begin{align}
&  M_D =v_D
P_R^*
\begin{pmatrix}
0  & 0 & a_D  \\
0&b_D & c_D  (2 {\rm Im} \tau)^{6} |Y_{\bf 1'}^{(12)}|\\
d_D& e_D  (2 {\rm Im} \tau)^{2} |Y_{\bf 1}^{(4)}|&
f_D (2 {\rm Im} \tau)^{8}(g_D Y_{\bf 1'{\rm A}}^{(16)}+
Y_{\bf 1'{\rm B}}^{(16)}) e^{-i (\varphi_4+\varphi_{12})}\end{pmatrix} P_L\,,
\nonumber\\
&M_U =v_U
P_R^*
\begin{pmatrix}
0  & 0 & a_U  \\
0&b_U & c_U (2 {\rm Im} \tau)^{6} |Y_{\bf 1'}^{(12)}|\\
d_U& e_U (2 {\rm Im} \tau)^{2} |Y_{\bf 1}^{(4)}|&
f_U (2 {\rm Im} \tau)^{8} (g_U Y_{\bf 1'{\rm A}}^{(16)}+
Y_{\bf 1'{\rm B}}^{(16)}) e^{-i (\varphi_4+\varphi_{12})}\end{pmatrix}
P_L
\,.
\label{model-1-PH}
\end{align}
%
The phase matrices $P_R$ and $P_L$ are given by:
\begin{align}
P_R=
\begin{pmatrix}
e^{i \varphi_{12}}  & 0 & 0  \\
0&1 & 0\\
0&0&e^{-i\varphi_4} \end{pmatrix}\,, \qquad
P_L=
\begin{pmatrix}
e^{-i \varphi_4}  & 0 & 0  \\
0&1 & 0\\
0&0&e^{i\varphi_{12}} \end{pmatrix}
\,,
\label{PH1}
\end{align}
%
where 
\begin{align}
\varphi_4\equiv \arg \,Y_{\bf 1}^{(4)}\,,\qquad 
\varphi_{12}\equiv \arg \,Y_{\bf 1'}^{(12)}\,.
\label{Arg-PH1}
\end{align}
%
The matrix $P_R$ does not appear in the CKM mixing matrix.
Since $P_L$ is common in down-type quark and up-type quark mass matrices,
the phase matrix $P_L$ is cancelled out in  the CKM matrix due to  
$P_L^* P_L=1$. Thus, the CP violation originates from the phases
of the $(3,3)$ elements of $M_D$ and $M_U$ 
in Eq.\,\eqref{model-1-PH}:
  \begin{align}
  \arg \ [ (g_D Y_{\bf 1'{\rm A}}^{(16)}+Y_{\bf 1'{\rm B}}^{(16)}) 
  \ e^{-i (\varphi_4+\varphi_{12})}] \equiv \Phi^{(1)}_D\,,\qquad 
   \arg \ [(g_U Y_{\bf 1'{\rm A}}^{(16)}+Y_{\bf 1'{\rm B}}^{(16)}) 
  \ e^{-i (\varphi_4+\varphi_{12})}] \equiv \Phi^{(1)}_U\,.
  \label{phase-1}
   \end{align}
%
These phases are fixed by the values of the VEV of modulus $\tau$ and of
the real parameters $g_D$ and $g_U$.
Due to the two ${\bf 1'}$ singlet  weight 16 modular forms, 
non-trivial CP-violating phases appear in down-type and up-type
quark mass matrices as well as in the CKM matrix.

In  Appendix \ref{Appen-texture-1} we show that the mass matrices 
$M_Q$, $Q=D,U$, given in Eq.\,\eqref{model-1-PH}, which correspond to the 
first (upper) matrix of Yukawa couplings in 
Eq.\,\eqref{textures}, are completely consistent 
with the observed quark masses and CKM quark mixing and CP violation 
when their elements have the numerical values 
reported in Eq.\,\eqref{texture-N1} of  Appendix \ref{Appen-texture-1}
and the CPV phases in  the (3,3) elements in $M_D$ and $M_U$ 
have the values $24.36^\circ$ and $-\,150.26^\circ$, respectively.
This implies that the quark matrices in Eq.\,\eqref{model-1-PH}, 
obtained using the $A_4$ modular invariance, 
could be consistent with the data on quark masses, mixing and CP 
violation if the phases $\Phi^{(1)}_D$ and  $\Phi^{(1)}_U$ 
in Eq.\,\eqref{phase-1} have the indicated values. 
If we will be able to adjust the VEV of the modulus $\tau$ 
and the values of the real constants $g_D$ and $g_U$ 
so that the phases $\Phi^{(1)}_D$ and  $\Phi^{(1)}_U$ 
get these values, we will have  viable 
quark mass matrices since 
the constants $a_D,\, b_D,\, c_D,\, d_D,\, e_D,\,f_D$ and 
$a_U,\, b_U,\, c_U,\, d_U,\, e_U,\,f_U$ 
can be used to reproduce the numerical 
values in the matrices in Eq.\,\eqref{texture-N1}.
These numerical values are derived by fitting the quark masses, 
the CKM matrix elements and the Jarlskog rephasing invariant $J_{CP}$ 
quoted in Eqs.\,\eqref{Datamass}, \eqref{DataCKM-GUT} and 
\eqref{JCPGUT} in Appendix \ref{Appen-inputs}.

We plot in Fig.\,\ref{tau-(1)} the values of the modulus $\tau$
\footnote{By ``values of the modulus $\tau$'' here and in what follows 
we mean ``values of the VEV of the modulus $\tau$''.
}, 
which allow to reproduce the quoted numerical values 
of the CPV phases $\Phi^{(1)}_D = 24.36^\circ$ 
and  $\Phi^{(1)}_U = -\,150.26^\circ$ within $1\%$ deviation 
by scanning $g_D$ and  $g_U$ in the ranges of 
$|g_D|,|g_U|=0-0.5$ (blue points), $|g_D|,|g_U|=0.5-2$ (red points)
and $|g_D|,|g_U|=2-10$ (magenta points). 
These ranges are chosen in order to illustrate the 
$g_D$ and  $g_U$ dependences. 
As is indicated by the figure, there is a viable   
region of values of $\tau$ up to   ${\rm Im}\,\tau\simeq 2$. 
We find a region near the fixed points $\tau=\omega$ 
that is also  viable.
Indeed, in subsection \ref{sec:fits-1} 
we easily obtain an example of a parameter set 
near the fixed points $\tau=\omega$ 
which provides a good description of the quark data.
There is scarcely a viable point close to $\tau=i$.

In Fig.\,\ref{gd-(1)} we show the corresponding 
region in the $g_D - {\rm Im}\,\tau$ plane 
for values of $|g_D|,|g_U|<10$. 
The imaginary part of $\tau$
can have relatively large values 
${\rm Im}\,\tau \cong (1.5 - 2.0)$ 
in the region of  $|g_D| < 0.5$, 
the maximal value ${\rm Im}\,\tau \cong 2.0$  
being reached for $|g_D| \ll 1$.
We do not show  the analogous plot in the $g_U - {\rm Im}\,\tau$ plane
since it is almost the same as the one in Fig.\,\ref{gd-(1)}.

We will present in subsection \ref{sec:fits-1} the results of the fits of the 
the quark masses, CKM elements and the Jarlskog rephasing invariant 
$J_{CP}$ quoted in Eqs.\,\eqref{Datamass}, \eqref{DataCKM-GUT} and 
\eqref{JCPGUT} of Appendix \ref{Appen-inputs}.

%
\begin{figure}[H]
	\begin{tabular}{ccc}
		\begin{minipage}{0.47\hsize}
				\includegraphics[{width=\linewidth}]{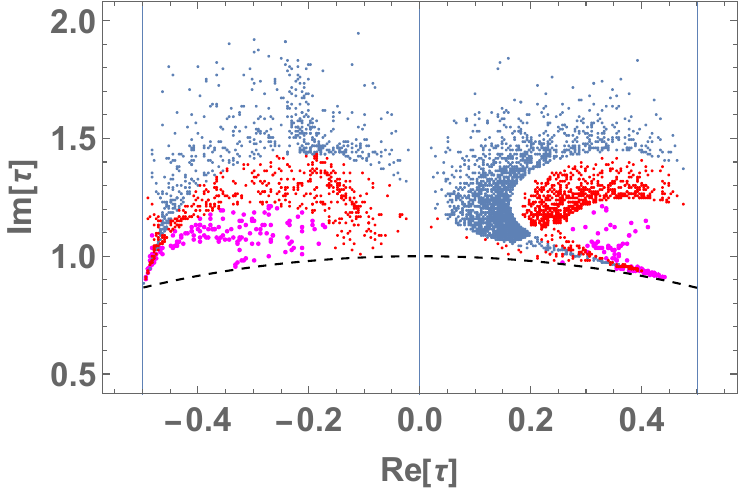} 
\vspace{-7mm}
\caption{The region in ${\rm Re}\,\tau$-${\rm Im}\,\tau$
plane consistent with with observed CP phase of CKM matrix in model (1).
Blue, red and magenta points correspond to
$|g_D|,|g_U|=0$-$0.5$, $0.5$-$2$ and  $2$-$10$, respectively.}
\label{tau-(1)}
\end{minipage}
\hskip 0.7 cm
\begin{minipage}{0.47\hsize}
 \includegraphics[width=\linewidth]{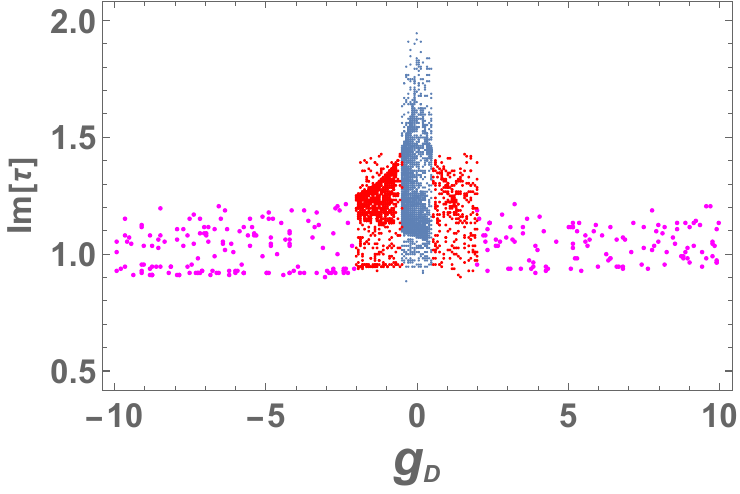}
\caption{The region in $g_D$-${\rm Im}\,\tau$
plane consistent with observed CP phase of CKM matrix in model (1).
Blue, red and magenta points correspond to
$|g_D|,|g_U|=0$-$0.5$, $0.5$-$2$ and  $2$-$10$, respectively.
}
\label{gd-(1)}
\end{minipage}
\end{tabular}
\end{figure}
%

Let us compare the allowed region obtained by us
in the $A_4$ modular model with quark mass matrices  given 
in Eq.\,(\ref{model-1-PH})
with that in the model proposed in Ref. \cite{Feruglio:2023uof}, 
in which modular forms of level $N=1$ modular symmetry 
are used. These modular forms are  SL$(2,\mathbb{Z})$ 
singlets - the Eisenstein series with weight $4$, $8$ and $12$, 
$E_4$, $E_8$ and $E_{12 A}$, $E_{12 B}$. 
They coincide with the singlet ${\bf 1}$ 
modular forms of the $A_4$ modular group of the same weights:   
$E_4 \equiv Y^{(4)}_{\bf 1}$, $E_8 \equiv Y^{(8)}_{\bf 1} = E^2_4$, 
 $E_{12 A} = E^3_4 \equiv Y^{(12)}_{{\bf 1}A}$, $E_{12 B} = 
E^2_6 -  E^3_4 \equiv  Y^{(12)}_{{\bf 1}B}$, 
where $E_6 \equiv Y^{(6)}_{\bf 1}$ and $Y^{(4)}_{\bf 1}$, 
 $Y^{(6)}_{\bf 1}$, 
$Y^{(8)}_{\bf 1}$, $Y^{(12)}_{{\bf 1}A}$ and $ Y^{(12)}_{{\bf 1}B}$ 
are given in Appendix \ref{modular-forms}. 
The parameters $g_D$ and $g_U$ appear in the 
quark mass matrices of the model in the same way they appear
in the Yukawa matrices in  Eq.\,\eqref{Ymodel-1}:
\begin{align}
 M_Q = v_Q\,\begin{pmatrix}
  0  & 0 & a_Q  \\
  0&b_Q & c_Q  (2 {\rm Im} \tau)^{4} Y_{\bf 1}^{(8)}\\
  d_Q& e_Q  (2 {\rm Im} \tau)^{2}  Y_{\bf 1}^{(4)}&
  f_Q (2 {\rm Im} \tau)^{6}
(g_Q Y_{\bf 1{\rm A}}^{(12)}+Y_{\bf 1 {\rm B}}^{(12)}) \end{pmatrix}_{RL}\,,
\quad Q=D,U\,.
  \label{Ymodel-2}
  \end{align}
 %
The results of this analysis performed by us 
are shown in Fig.\,\ref{tau-Eisen} and  Fig.\,\ref{gd-Eisen}.
The distributions of $\tau$ and $g_D$ are almost the same as
those shown in  Figs.\,\ref{tau-(1)}  and \ref{gd-(1)}
because the textures of $M_D$ and $M_U$ are the same in the two models.
We will see later that the distributions  of $\tau$ and $g_D$
depend significantly on the textures of the quark mass matrices.
\begin{figure}[H]
	\begin{tabular}{ccc}
		\begin{minipage}{0.47\hsize}
 \includegraphics[width=\linewidth]{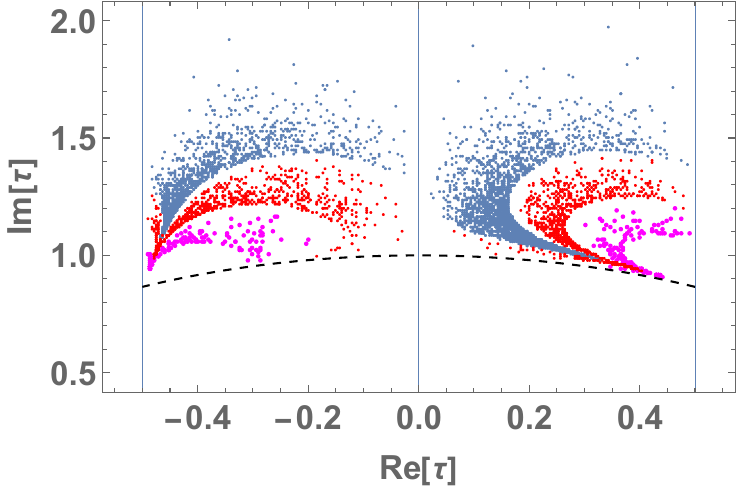}
\vspace{-7mm}
\caption{The region in ${\rm Re}\,\tau$-${\rm Im}\,\tau$ 
plane consistent with the observed CPV phase of CKM matrix in the 
case of  $N=1$ modular symmetry of the quark mass matrices, 
where Eisenstein series are used as modular forms 
\cite{Feruglio:2023uof}.
Blue, red and magenta points correspond to
$|g_D|,|g_U|=0$-$0.5$, $0.5$-$2$ and  $2$-$10$, respectively.
See the text for further details.}
\label{tau-Eisen}
\end{minipage}
\hskip 0.7 cm
\begin{minipage}{0.47\hsize}
  \includegraphics[width=\linewidth]{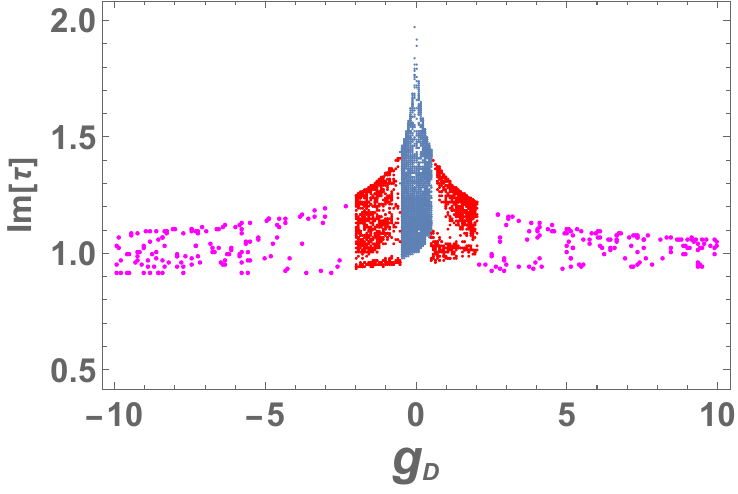}
\caption{The region in $g_D$-${\rm Im}\,\tau$
plane consistent with observed CP phase of CKM matrix
in the case of  $N=1$.
Blue, red and magenta points correspond to
$|g_D|,|g_U|=0$-$0.5$, $0.5$-$2$ and  $2$-$10$, respectively.}
\label{gd-Eisen}
\end{minipage}
\end{tabular}
\end{figure}

%
\subsection{Quark mass matrix of model (2) and its CPV phase structure} 
%
%
We  present the second  model (2), in which the assignments 
of  the weights for the relevant chiral superfields are as follows:
\begin{itemize}
	\item{quark doublet (left-handed) $Q_1 =(d,u)_L,Q_2=(s,c)_L,Q_3=(b,t)_L$: $A_4$ singlets
		$(1,\,1,\,1'')$ with weight $(-4,-8,8)$.}
	\item{quark singlets (righ-handed) $(d^c,s^c,b^c)$ and
		$(u^c,c^c,t^c)$
		: $A_4$ singlets $(1',\,1,\,1)$ with weight $(-8,4,8)$.}
	\item{ Higgs fields of down-type and up-type quark sectors $H_{U,D}$: $A_4$ singlet $1$ with weight 0.}
\end{itemize}
These assignments satisfy the weight condition of Eq.\,\eqref{weight-condition}. Those are summarized in Table \ref{tab:model-2}.
 \begin{table}[H]
 	\begin{center}
 		\renewcommand{\arraystretch}{1.1}
 		\begin{tabular}{|c|c|c|c|c|} \hline
 			& $(d,u)_L,(s,c)_L,(b,t)_L$ & $(d^c,s^c,b^c),\,(u^c,c^c,t^c)$ &  $H_U$ & $H_D$ \\ \hline
 			$SU(2)$ & 2 & 1  & 2 & 2 \\
 			$A_4$ & $(1,\,1\,,1'')$ & $(1',\,1\,,1)$ & $1$ & $1$ \\
 			$k$ &$ (-4,\ -8, 8)$  &$(-8,4,\ 8)$  & 0 & 0 \\ \hline
 		\end{tabular}
 	\end{center}
 	\caption{Assignments of $A_4$ representations and weights
 		 in  model (2).}
 	\label{tab:model-2}
 \end{table}
%
 The mass matrices of the down-type and up-type quarks read:
\begin{align}
 M_Q =v_Q
 \begin{pmatrix}
 0  & 0 & {a_Q} \\
 {b_Q}&0 & {c_Q}  (2 {\rm Im} \tau)^{6} Y_{\bf 1'}^{(12)}\\
 {e_Q}  (2 {\rm Im} \tau)^{2} Y_{\bf 1}^{(4)}& {d_Q}&
 {f_Q} (2 {\rm Im} \tau)^{8}(g_Q Y_{\bf 1'_{\rm A}}^{(16)}+
 Y_{\bf 1'_{\rm B}}^{(16)}) \end{pmatrix}_{RL}\,,
 \qquad Q=D, U \ .
 \label{model-2}
 \end{align}
%
 The CP violating phases come from the modulus $\tau$ in modular forms
 of  $Y_{\bf 1}^{(4)}$,  $Y_{\bf 1'}^{(12)}$,
 $Y_{\bf 1'_{\rm A}}^{(16)}$ and $Y_{\bf 1'_{\rm B}}^{(16)}$.
 The phases of  $Y_{\bf 1}^{(4)}$ and  $Y_{\bf 1'}^{(12)}$
 in the $(2,3)$ and $(3,1)$ elements can be 
factorised as follows:
\begin{align}
 M_Q =v_Q P_R^*
 \begin{pmatrix}
 0  & 0 & {a_Q} \\
 {b_Q}&0 & {c_Q}  (2 {\rm Im} \tau)^{6} |Y_{\bf 1'}^{(12)}|\\
 {e_Q}  (2 {\rm Im} \tau)^{2} |Y_{\bf 1}^{(4)}|& {d_Q}&
 {f_Q} (2 {\rm Im} \tau)^{8}(g_Q Y_{\bf 1'_{\rm A}}^{(16)}+
 Y_{\bf 1'_{\rm B}}^{(16)})e^{-i (\varphi_4+\varphi_{12})} \end{pmatrix}_{RL} 
\hskip -0.3cm P_L\,,
 \ \  Q=D, U \ .
 \label{model-2-PH}
 \end{align}
%
The phase matrices $P_R$ and $P_L$ are given by:
\begin{align}
 P_R=
 \begin{pmatrix}
 e^{i \varphi_{12}}  & 0 & 0  \\
 0&1 & 0\\
 0&0&e^{-i\varphi_4} \end{pmatrix}\,, \qquad
 P_L=
 \begin{pmatrix}
 1 & 0 & 0  \\
 0&e^{-i \varphi_{4}}  & 0\\
 0&0&e^{i\varphi_{12}} \end{pmatrix}
 \,,
 \label{PH2}
 \end{align}
%
where $ \varphi_4$ and $ \varphi_{12}$ are defined in Eq.\eqref{Arg-PH1}.
As in the model (1), the matrix $P_R$ does not contribute to the 
CKM mixing matrix, while  $P_L$, being common to down-type quark and up-type 
quark mass matrices,  is cancelled out in the CKM matrix.
 The CP violation is generated by the phases
of the $(3,3)$ elements of $M_D$ and $M_U$  in Eq.\,\eqref{model-2-PH},
 \begin{align}
 \arg \ [(g_D Y_{\bf 1{\rm A}}^{(16)}+Y_{\bf 1{\rm B}}^{(16)}) 
 \ e^{-i (\varphi_4+\varphi_{12})}] \equiv \Phi^{(2)}_D\,,\qquad 
 \arg \ [(g_U Y_{\bf 1{\rm A}}^{(16)}+Y_{\bf 1{\rm B}}^{(16)}) 
 \ e^{-i (\varphi_4+\varphi_{12})}] \equiv \Phi^{(2)}_U \,,
 \label{phase-2}
 \end{align}
%
which are determined by the VEV of the modulus $\tau$ and 
the real parameters $g_D$ and $g_U$.

 The analysis which follows is analogous to that performed for the 
model (1) in the preceding section.
The second (middle) matrix of  Yukawa couplings in Eq.\,\eqref{textures} 
generates the quark mass matrices reported in Eq.\,\eqref{texture-2} 
of Appendix \ref{Appen-texture-2}, which have the same structure as the 
the mass matrices in Eq.\,\eqref{model-2-PH},
obtained by employing the $A_4$ modular symmetry.
It is shown in  \ref{Appen-texture-2} that 
these mass matrices successfully describe the data on the  
quark masses, mixing and CP violation 
if their elements have the numerical values 
given in  Eq.\,\eqref{texture-N2}
and the CPV phases in  the (3,3) elements in $M_D$ and $M_U$ 
posses the values $104.84^\circ$ and $56.63^\circ$, respectively.
Reproducing the real numerical values of the elements of 
$M_D$ and $M_U$ reported in Eq.\,\eqref{texture-N2} using 
the constants $a_D,\, b_D,\, c_D,\, d_D,\, e_D,\,f_D$ and 
$a_U,\, b_U,\, c_U,\, d_U,\, e_U,\,f_U,$ 
does not pose a problems.
Thus, if we find  values of the VEV of $\tau$ and of the real constants 
$g_D$ and $g_U$ that generate CPV phases 
$\Phi^{(2)}_D = 104.84^\circ$ and  $\Phi^{(2)}_U = 56.63^\circ$,  
then the $A_4$ modular matrices $M_D$ and $M_U$ in 
Eq.\,\eqref{model-2-PH} will be phenomenologically viable for the 
so found values of the modulus VEV and $g_D$ and $g_U$.

We have performed the relevant analysis and  
show  in Fig.\,\ref{tau-(2)} the region of values of the VEV
of $\tau$ for which the phases $\Phi^{(2)}_D = 104.84^\circ$ and  
$\Phi^{(2)}_U = 56.63^\circ$ are reproduced 
within $1\%$ deviation
by scanning $g_D$ and  $g_U$ in the ranges of 
 	$|g_D|,|g_U|=0-0.5$ (blue points), $|g_D|,|g_U|=0.5-2$ (red points)
 	and $|g_D|,|g_U|=2-10$ (magenta points).
The allowed points of $\tau$  are considerably reduced
as compared with those  in  Fig.\,\ref{tau-(1)}.
However, there is still a viable   
region of values of $\tau$ close to  ${\rm Im}\,\tau= 1.9$. 
We also find a region near the fixed points $\tau=\omega$ 
that is also  viable.
Indeed, in subsection \ref{sec:fits-2} 
we present  an example of a parameter set 
near the fixed points $\tau=\omega$ as well as the ones 
close to $\tau=i$ and at the large ${\rm Im}\,\tau$. 

In Fig.\,\ref{gd-(2)} we show the corresponding 
region in the $g_D - {\rm Im}\,\tau$ plane 
for values of $|g_D|,|g_U|<10$. 
The imaginary part of $\tau$
reaches the value of $1.9$ for  $|g_D|\ll 1$.
The analogous plot of $g_U - {\rm Im}\,\tau$ plane
is almost the same one as the one  in Fig.\,\ref{gd-(2)} 
and we do not show it here.

We will present in subsection \ref{sec:fits-2} the results of the fits of the 
the quark masses, CKM elements and the Jarlskog rephasing invariant 
$J_{CP}$ quoted in Eqs.\,\eqref{Datamass}, \eqref{DataCKM-GUT} and 
\eqref{JCPGUT} of Appendix \ref{Appen-inputs}.

 \begin{figure}[H]
 	\begin{tabular}{ccc}
 		\begin{minipage}{0.47\hsize}
 \includegraphics[width=\linewidth]{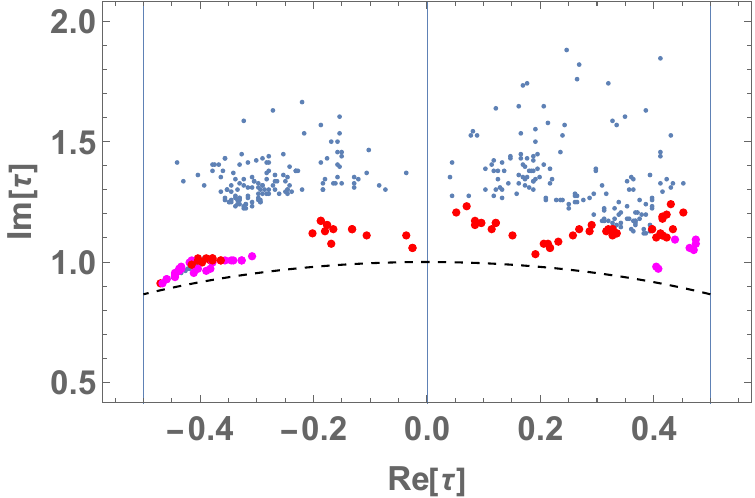} \vspace{-7mm}
\caption{The region in ${\rm Re}\,\tau$-${\rm Im}\,\tau$
 plane consistent with with observed CP phase of CKM matrix in model (2).
 Blue, red and magenta points correspond to
 $|g_D|,|g_U|=0$-$0.5$, $0.5$-$2$ and  $2$-$10$, respectively. }
 \label{tau-(2)}
 \end{minipage}
 \hskip 0.7 cm
\begin{minipage}{0.47\hsize}
  \includegraphics[width=\linewidth]{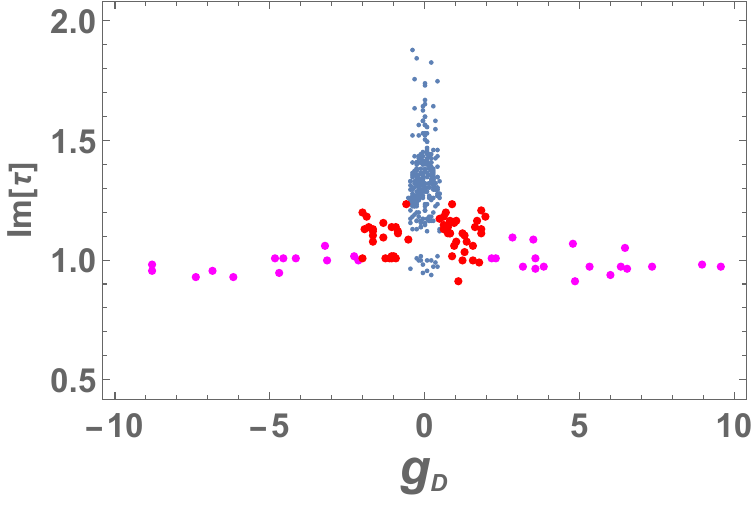}
 \caption{The region in $g_D$-${\rm Im}\,\tau$
 plane consistent with observed CP phase of CKM matrix
 in model (2).
Blue, red and magenta points correspond to
$|g_D|,|g_U|=0$-$0.5$, $0.5$-$2$ and  $2$-$10$, respectively.}
 \label{gd-(2)}
 \end{minipage}
\end{tabular}
 \end{figure}
%

%
 \subsection{Quark mass matrix of model (3) and its CPV phase structure}
%

For the model (3), the assignments of  the weights 
for the relevant chiral superfields is given  as follows:
\begin{itemize}
	\item{quark doublet (left-handed) $Q_1 =(d,u)_L,Q_2=(s,c)_L,Q_3=(b,t)_L$: $A_4$ singlets
		$(1'',\,1,\,1)$ with weight $(8,-4,-8)$.}
	\item{quark singlets (righ-handed) $(d^c,s^c,b^c)$ and
		$(u^c,c^c,t^c)$
		: $A_4$ singlets $(1',\,1,\,1)$ with weight $(-8,4,8)$.}
	\item{ Higgs fields of down-type and up-type quark sectors $H_{U,D}$: $A_4$ singlet $1$ with weight 0.}
\end{itemize}
These assignments satisfy the weight condition of Eq.\eqref{weight-condition}. Those are summarized in Table \ref{tab:model-3}.
 \begin{table}[H]
 	\begin{center}
 		\renewcommand{\arraystretch}{1.1}
 		\begin{tabular}{|c|c|c|c|c|} \hline
 			& $(d,u)_L,(s,c)_L,(b,t)_L$ & $(d^c,s^c,b^c),\,(u^c,c^c,t^c)$ &  $H_U$ & $H_D$ \\ \hline
 			$SU(2)$ & 2 & 1  & 2 & 2 \\
 			$A_4$ & $(1'',\,1\,,1)$ & $(1',\,1\,,1)$ & $1$ & $1$ \\
 			$k$ &$ (8,\ -4,\ -8)$  &$(-8,4,\ 8)$  & 0 & 0 \\ \hline
 		\end{tabular}
 	\end{center}
 	\caption{Assignments of $A_4$ representations and weights
 		 in  model (3).}
 	\label{tab:model-3}
 \end{table}
%
The mass matrices of the down-type and up-type quarks have the form:
\begin{align}
 M_Q =v_Q
 \begin{pmatrix}
 {a_Q}  & 0 &0 \\
 {c_Q}  (2 {\rm Im} \tau)^{6} Y_{\bf 1'}^{(12)}&{b_Q} & 0 \\
 {f_Q} (2 {\rm Im} \tau)^{8}(g_Q Y_{\bf 1'{\rm A}}^{(16)}+
 Y_{\bf 1'{\rm B}}^{(16)})& {e_Q}  (2 {\rm Im} \tau)^{2} Y_{\bf 1}^{(4)}&{d_Q}
 \end{pmatrix}_{RL}\,,
 \qquad Q=D, U \ .
 \label{model-3-PH}
 \end{align}
%

Since the number of parameters are 14,  four parameters are
redundant for reproducing the quark masses 
and CKM elements (10 observables).
In this case, following Ref.\cite{Tanimoto:2016rqy},
we set  $c_U=e_U=f_U=0$ and we investigate whether 
one can describe successfully the quark data 
with this additional phenomenological assumption. 
As a consequence of setting the constants 
$c_U$, $e_U$ and $f_U$ to zero the up-type quark mass matrix is diagonal.

The CP violating phases originate from the modulus $\tau$ and appear 
in the down-type quark mass matrix $M_D$ via 
the modular forms  $Y_{\bf 1}^{(4)}$,  $Y_{\bf 1'}^{(12)}$,
$Y_{\bf 1'_{\rm A}}^{(16)}$ and $Y_{\bf 1'_{\rm B}}^{(16)}$.
The phases of  $Y_{\bf 1}^{(4)}$ and  $Y_{\bf 1'}^{(12)}$
in the $(2,3)$ and $(3,1)$ elements can be factored out as follows:
 \begin{align}
 M_D =v_D P_R^*
 \begin{pmatrix}
 {a_D}  & 0 & 0 \\
 {c_D}  (2 {\rm Im} \tau)^{6} |Y_{\bf 1'}^{(12)}|&{b_D} & 0\\
 {f_D} (2 {\rm Im} \tau)^{8}(g_D Y_{\bf 1'_{\rm A}}^{(16)}+
 Y_{\bf 1'_{\rm B}}^{(16)})e^{-i (\varphi_4+\varphi_{12})}&
 {e_D}  (2 {\rm Im} \tau)^{2} |Y_{\bf 1}^{(4)}|& {d_D}
 \end{pmatrix}_{RL} P_L\,.
 \label{model-3-PHD}
 \end{align}
%
 The phase matrices $P_R$ and $P_L$ are 
\begin{align}
 P_R=P_L=
 \begin{pmatrix}
 1 & 0 & 0  \\
 0&e^{-i \varphi_{12}}  & 0\\
 0&0&e^{-i(\varphi_{4}+\varphi_{12})} \end{pmatrix}
 \,,
 \label{(2)-PH}
 \end{align}
%
where $ \varphi_4$ and $ \varphi_{12}$ are given in Eq.\,\eqref{Arg-PH1}.

On the other hand,  the  up-type quark mass matrix is diagonal.
Therefore,  $P_L$ and $P_R$ do not contribute to the CKM matrix.
Thus, the CP violation is generated by the phase
 of the $(3,1)$ element of $M_D$
 in Eq.\,\eqref{model-3-PH},
\begin{align}
 \arg \ [(g_D Y_{\bf 1'{\rm A}}^{(16)}+Y_{\bf 1'{\rm B}}^{(16)}) 
 \ e^{-i (\varphi_4+\varphi_{12})} ]\equiv \Phi^{(3)}_D\,,
 \label{phase-3}
 \end{align}
%
 which is determined by the VEV of $\tau$ and the real parameter $g_D$.

As shown in Eq.\,\eqref{texture-3} of Appendix \ref{Appen-texture-3}, 
the quark mass matrices
 in Eq.\,\eqref{model-3-PH} could be completely consistent with observed 
masses and the CKM matrix if the phase of Eq.\,\eqref{phase-3} 
has the value $\Phi^{(3)}_D = 66.36^\circ$. 
By adjusting the value of the VEV of the modulus $\tau$ and of 
the real constant $g_D$ to get $\Phi^{(3)}_D = 66.36^\circ$,
we will obtain a phenomenologically viable down-quark mass matrix 
$M_D$. The real constants  $a_D,\, b_D,\, c_D,\, d_D,\, e_D,\,f_D$ 
can be used  to reproduce the real numerical values in 
$M_D$ in  Eq.\,\eqref{texture-N3}.

 We plot in Fig.\,\ref{tau-(3)} the values of the VEV of $\tau$ for which 
one can generate $\Phi^{(3)}_D = 66.36^\circ$
within $1\%$ deviation
by scanning $g_D$ and  $g_U$ again in the ranges of 
	$|g_D|,|g_U|=0-0.5$ (blue points), $|g_D|,|g_U|=0.5-2$ (red points)
	and $|g_D|,|g_U|=2-10$ (magenta points).
In the considered case it is possible to reproduce 
the observed CP violation in the quark sector even for 
$\tau$ close to $\omega$ and $\tau$ ``close'' to $i\infty$.
  Indeed, we present  examples of viable parameter sets
for  $\tau\simeq \omega$ and ${\rm Im}\, \tau\simeq 1.7$ in subsection \ref{sec:fits-3}.

 In Fig.\,\ref{gd-(3)}, we show the values of  
  ${\rm Im}\,\tau$ leading (within $1\%$ deviation) 
to the requisite value of $\Phi^{(3)}_D$ versus  $|g_D|<10$.
 The magnitude of ${\rm Im}\,\tau$ increases rapidly when $|g_D|$
  decreases towards $0$.
 \begin{figure}[H]
 	\begin{tabular}{ccc}
 		\begin{minipage}{0.47\hsize}
 \includegraphics[width=\linewidth]{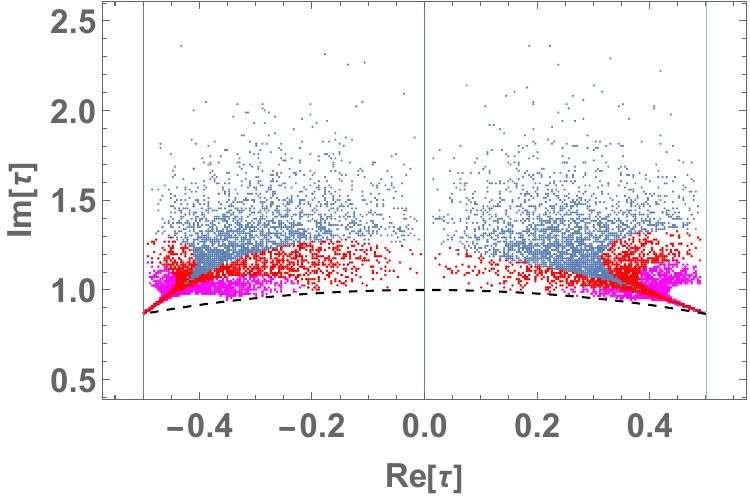} 
 			\caption{The region in ${\rm Re}\,\tau$-${\rm Im}\,\tau$
 				plane consistent with with observed CP phase of CKM matrix in model (3).
 			Blue, red and magenta points correspond to
 			$|g_D|,|g_U|=0$-$0.5$, $0.5$-$2$ and  $2$-$10$, respectively. }
 			\label{tau-(3)}
 		\end{minipage}
 		\hskip 0.7 cm
 		\phantom{=}
 		\begin{minipage}{0.46\hsize}
 	\includegraphics[width=\linewidth]{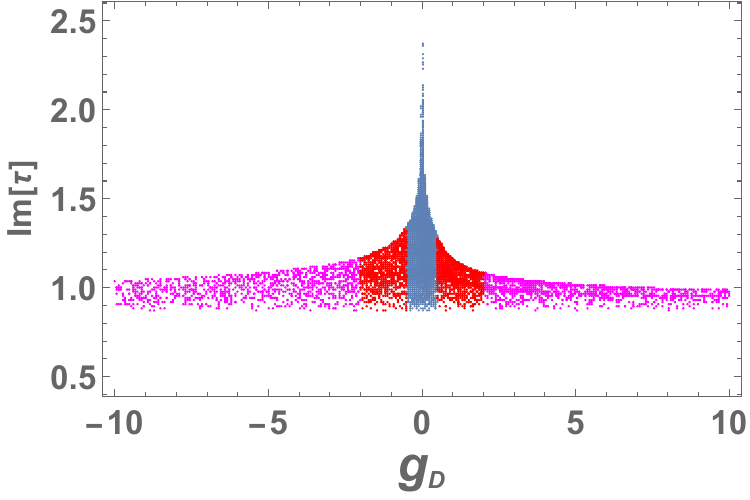}
 			\caption{The region in $g_D$-${\rm Im}\,\tau$
 				plane consistent with observed CP phase of CKM matrix
 				in model (3).
 			Blue, red and magenta points correspond to
 			$|g_D|,|g_U|=0$-$0.5$, $0.5$-$2$ and  $2$-$10$, respectively.}
 			\label{gd-(3)}
 		\end{minipage}
 	\end{tabular}
 \end{figure}
%

Let us emphasise that in 
the case of the discussed model (3) we have set 
the redundant parameters $c_U=e_U=f_U=0$ 
``by hand'' to get diagonal mass matrix for the up-type quarks. 
This set-up is not guaranteed within the framework of
the modular invariance approach without additional (symmetry) 
assumptions.

%
\section{Reproducing quark masses and CKM parameters}
\label{sec:fits}
%
%
We present next numerical examples of successfully reproducing the 
observed quark masses, the CKM mixing angles and  CPV phase 
in the cases of the three models considered by us.

%
\subsection{Fitting model (1)}
\label{sec:fits-1}
%
%

We present three examples of model (1) corresponding to the 
$\tau$ being relatively close to the fixed points $\tau=i$, $\omega$ and
$\infty$, respectively.
%
\subsubsection{$\tau$ close to $i$}
%
%
The first one is an  example of model (1), where
$\tau$ is rather close to $i$. 
This case is comparable to the example
 in Ref.\cite{Feruglio:2023uof}.
We show the numerical values of parameters 
obtained in the fit of  
the quark masses, CKM mixing angles, the CPV phase $\delta_{CP}$ and 
of the Jarlskog rephasing invariant $J_{CP}$ \cite{Jarlskog:1985ht}, 
as defined in \cite{ParticleDataGroup:2022pth} and  
quoted in Eqs.\,\eqref{Datamass}, \eqref{DataCKM-GUT} and 
\eqref{JCPGUT} of Appendix \ref{Appen-inputs}: 
\begin{align}
&\frac{b_D}{a_D}=-0.724,\quad \frac{c_D}{a_D}=0.0430,\quad
\frac{d_D}{a_D}=0.0864,\quad \frac{e_D}{a_D}=0.135,\quad 
\frac{f_D}{a_D}=0.00750,\nonumber\\
&\frac{b_U}{a_U}=0.972,\quad \frac{c_U}{a_U}=0.361,\quad
 \frac{d_U}{a_U}=0.631,\quad \frac{e_U}{a_U}=1.49,\quad 
\frac{f_U}{a_U}=0.154,\nonumber\\
&g_D=-0.749,\quad g_U=-1.10, \quad \tau=0.1811+ i\,1.1583, \quad N\sigma=0.52,
\label{parameters-(1)-1}
\end{align}
%
where $N\sigma$ denotes a measure of goodness of the fit.
By employing the sum of one-dimensional 
$\Delta\chi^2$ for eight   observable quantities
$m_d/m_b, \, m_s/m_b,\,m_u/m_t, \, m_c/m_t,\, |V_{us}|,\, |V_{cb}|,\, |V_{ub}|,\, \delta_{CP}$,
 it is defined as $N\sigma\equiv \sqrt{\Delta \chi^2}$.
%
The result of the fit of the quark mass ratios,
three CKM elements $|V_{us}|$,  $|V_{cb}|$,  $|V_{ub}|$,
the phase  $\delta_{CP}$ and of the $J_{CP}$ factor 
are collected in Table \ref{tab:output-(1)-1}.
\begin{table}[H]
	\small{
		\begin{center}
			\renewcommand{\arraystretch}{1.1}
			\begin{tabular}{|c|c|c|c|c|c|c|c|c|c|} \hline
				\rule[14pt]{0pt}{3pt}  
				& $\frac{m_s}{m_b}\hskip -1 mm\times\hskip -1 mm 10^2$ 
				& $\frac{m_d}{m_b}\hskip -1 mm\times\hskip -1 mm 10^4$& $\frac{m_c}{m_t}\hskip -1 mm\times\hskip -1 mm 10^3$&$\frac{m_u}{m_t}\hskip -1 mm\times\hskip -1 mm 10^6$&
				$|V_{us}|$ &$|V_{cb}|$ &$|V_{ub}|$&$|J_{\rm CP}|$& $\delta_{\rm CP}$
				\\
				\hline
				\rule[14pt]{0pt}{3pt}  
				Fit &$1.87$ & $8.84$
				& $2.80$ & $5.89$&
				$0.2251$ & $0.0401$ & $0.00352$ &
				$2.84\hskip -1 mm\times\hskip -1 mm 10^{-5}$&$66.8^\circ$
				\\ \hline
				\rule[14pt]{0pt}{3pt}
				Exp	 &$1.82$ & $9.21$ 
				& $2.80$& $ 5.39$ &
				$0.2250$ & $0.0400$ & $0.00353$ &$2.80\hskip -1 mm\times\hskip -1 mm 10^{-5}$&$66.2^\circ$\\
				$1\,\sigma$	&$\pm 0.10$ &$\pm 1.02$ & $\pm 0.12$& $\pm 1.68$ &$ \pm 0.0007$ &
				$ \pm 0.0008$ & $ \pm 0.00013$ &$^{+0.14}_{-0.12}\hskip -1 mm\times \hskip -1 mm 10^{-5}$&
				$^{+ 3.4^\circ}_{-3.6^\circ}$\\ \hline  
			\end{tabular}
		\end{center}
		\caption{Results of the fits of the quark mass ratios, 
			CKM mixing angles, $J_{\rm CP}$ and $\delta_{\rm CP}$. 'Exp' denotes the  values of the observables 
			at the GUT scale, including $1\sigma$ error.
		}
		\label{tab:output-(1)-1}
	}
\end{table}

  We comment on the values of parameters in Eq.\,\eqref{parameters-(1)-1}.
The parameters $a_Q$, $b_Q$, $d_Q$ ($Q=D,U$) are real constants that 
do not couple  to modular forms. On the other hand,  
$c_Q$, $e_Q$, $f_Q$ ($Q=D,U$) are constants multiplying the modular forms.
Since the normalizations of each of the three
modular forms 
present in the expressions for the quark mass matrices $M_D$ and $M_U$
are arbitrary, the magnitudes of the parameters  $a_Q$, $b_Q$, $d_Q$ 
and respectively those of  $c_Q$, $e_Q$, $f_Q$  
should be compared separately among  themselves.
The constants multiplying the modular forms are  
of the same order in magnitude:
 \begin{align}
 \left|\frac{c_D}{f_D}\right |\simeq 5.7\,,\qquad \left|\frac{e_D}{f_D}\right|\simeq 18\,,\qquad
 \left|\frac{c_U}{f_U}\right|\simeq 2.3\,,\qquad \left|\frac{e_U}{f_U}\right|\simeq 9.7\,.
 \end{align}
%
In what concerns the constants, $a_Q$, $b_Q$, $d_Q$, 
only the ration $|d_D/a_D|$ is somewhat smaller 
than the other ratios $|b_D/a_D|$, 
$|b_U/a_U|$ and $|d_U/a_U|$:
\begin{align}
 \left|\frac{b_D}{a_D}\right |\simeq 0.72\,,\qquad \left|\frac{d_D}{a_D}\right|\simeq 0.086\,,\qquad
 \left|\frac{b_U}{a_U}\right|\simeq 0.97\,,\qquad \left|\frac{d_U}{a_U}\right|\simeq 0.63\,.
 \end{align}

%
\subsubsection{$\tau$ close to $\omega$}
%
%
We present the second example of successful fit of the quark data, 
in which the VEV of the modulus $\tau$ is close to the fixed point 
$\omega = -\,0.5 + i\,\sqrt{3}/2$ (the left cusp).
The numerical values of parameters read: 
\begin{align}
&\frac{b_D}{a_D}=0.895,\quad \frac{c_D}{a_D}=0.816,\quad\frac{d_D}{a_D}=0.0830,\quad
 \frac{e_D}{a_D}=0.524,\quad 
\frac{f_D}{a_D}=0.00973,\nonumber\\
&\frac{b_U}{a_U}=1.26,\quad \frac{c_U}{a_U}=5.90,\quad \frac{d_U}{a_U}=0.667, \quad
 \frac{e_U}{a_U}=7.53,\quad 
\frac{f_U}{a_U}=0.264,\nonumber\\
&g_D=-0.809,\ \ g_U=3.33, \ \ \tau=-0.4900+ i\,0.9493, \ \ \tau-\omega=0.0093+ i\,0.0832\,, \  N\sigma=0.87.
\label{parameters-(1)-2}
\end{align}
%

In Table \ref{tab:output-(1)-2}
we present the results of the fit of the quark mass ratios,  
the three CKM elements $|V_{us}|$,  $|V_{cb}|$,  $|V_{ub}|$, of the 
CPV phase  $\delta_{CP}$ and of the  $J_{CP}$ factor.
\begin{table}[H]
	\small{
		\begin{center}
			\renewcommand{\arraystretch}{1.1}
			\begin{tabular}{|c|c|c|c|c|c|c|c|c|c|} \hline
				\rule[14pt]{0pt}{3pt}  
				& $\frac{m_s}{m_b}\hskip -1 mm\times\hskip -1 mm 10^2$ 
				& $\frac{m_d}{m_b}\hskip -1 mm\times\hskip -1 mm 10^4$& $\frac{m_c}{m_t}\hskip -1 mm\times\hskip -1 mm 10^3$&$\frac{m_u}{m_t}\hskip -1 mm\times\hskip -1 mm 10^6$&
				$|V_{us}|$ &$|V_{cb}|$ &$|V_{ub}|$&$|J_{\rm CP}|$& $\delta_{\rm CP}$
				\\
				\hline
				\rule[14pt]{0pt}{3pt}  
				Fit &$1.88$ & $8.73$
				& $2.85$ & $5.57$&
				$0.2252$ & $0.0398$ & $0.00349$ &
				$2.80\hskip -1 mm\times\hskip -1 mm 10^{-5}$&$66.9^\circ$
				\\ \hline
				\rule[14pt]{0pt}{3pt}
				Exp	 &$1.82$ & $9.21$ 
				& $2.80$& $ 5.39$ &
				$0.2250$ & $0.0400$ & $0.00353$ &$2.80\hskip -1 mm\times\hskip -1 mm 10^{-5}$&$66.2^\circ$\\
				$1\,\sigma$	&$\pm 0.10$ &$\pm 1.02$ & $\pm 0.12$& $\pm 1.68$ &$ \pm 0.0007$ &
				$ \pm 0.0008$ & $ \pm 0.00013$ &$^{+0.14}_{-0.12}\hskip -1 mm\times \hskip -1 mm 10^{-5}$&
				$^{+ 3.4^\circ}_{-3.6^\circ}$\\ \hline  
			\end{tabular}
		\end{center}
		\caption{Results of the fits of the quark mass ratios, 
			CKM mixing angles, $J_{\rm CP}$ and $\delta_{\rm CP}$. 'Exp' denotes the  values of the observables 
			at the GUT scale, including $1\sigma$ error.
		}
		\label{tab:output-(1)-2}
	}
\end{table}
%

The ratios of the constants multiplying the modular forms 
are of the same order of magnitude:
\begin{align}
\left|\frac{c_D}{f_D}\right |\simeq 83.8\,,\qquad \left|\frac{e_D}{f_D}\right|\simeq 53.9\,,\qquad
\left|\frac{c_U}{f_U}\right|\simeq 22.3\,,\qquad \left|\frac{e_U}{f_U}\right|\simeq 28.5\,.
\end{align}
%
In what concerns the other constants, 
only the ratio of $|d_D/a_D|$ is again somewhat smaller than the 
other ratios:
\begin{align}
\left|\frac{b_D}{a_D}\right |\simeq 0.90\,,\qquad \left|\frac{d_D}{a_D}\right|\simeq 0.083\,,\qquad
\left|\frac{b_U}{a_U}\right|\simeq 1.26\,,\qquad \left|\frac{d_U}{a_U}\right|\simeq 0.67
\,.
\end{align}
%
%
\subsubsection{$\tau$ ``close'' to $i\infty$}
%

We present the third example of successful fit of quark data, 
in which the VEV of the modulus $\tau$ has a relatively large 
imaginary part ``close'' to the fixed point 
$i\infty$.
The numerical values of parameters read: 
\begin{align}
&\frac{b_D}{a_D}=-0.469,\quad\frac{c_D}{a_D}=0.0138,\quad \frac{d_D}{a_D}=0.0581,\quad
 \frac{e_D}{a_D}=0.0296,\quad 
\frac{f_D}{a_D}=0.0497,\nonumber\\
&\frac{b_U}{a_U}=1.585,\quad \frac{c_U}{a_U}=0.0867,\quad\frac{d_U}{a_U}=2.504, \quad
 \frac{e_U}{a_U}=0.601,\quad 
\frac{f_U}{a_U}=4.282,\nonumber\\
&g_D=0.000130,\ \ g_U=-0.00316, \ \ \tau=0.1097+ i\,1.8213, \ \  \  N\sigma=1.31.
\label{parameters-(1)-3}
\end{align}
%

In Table \ref{tab:output-(1)-3}
we present the results of the fit of the quark mass ratios,  
the three CKM elements $|V_{us}|$,  $|V_{cb}|$,  $|V_{ub}|$, of the 
CPV phase  $\delta_{CP}$ and of the  $J_{CP}$ factor.
\begin{table}[H]
	\small{
		\begin{center}
			\renewcommand{\arraystretch}{1.1}
			\begin{tabular}{|c|c|c|c|c|c|c|c|c|c|} \hline
				\rule[14pt]{0pt}{3pt}  
				& $\frac{m_s}{m_b}\hskip -1 mm\times\hskip -1 mm 10^2$ 
				& $\frac{m_d}{m_b}\hskip -1 mm\times\hskip -1 mm 10^4$& $\frac{m_c}{m_t}\hskip -1 mm\times\hskip -1 mm 10^3$&$\frac{m_u}{m_t}\hskip -1 mm\times\hskip -1 mm 10^6$&
				$|V_{us}|$ &$|V_{cb}|$ &$|V_{ub}|$&$|J_{\rm CP}|$& $\delta_{\rm CP}$
				\\
				\hline
				\rule[14pt]{0pt}{3pt}  
				Fit &$1.88$ & $9.71$
				& $2.82$ & $3.44$&
				$0.2248$ & $0.0401$ & $0.00353$ &
				$2.84\hskip -1 mm\times\hskip -1 mm 10^{-5}$&$66.6^\circ$
				\\ \hline
				\rule[14pt]{0pt}{3pt}
				Exp	 &$1.88$ & $8.72$ 
				& $2.80$& $ 5.39$ &
				$0.2250$ & $0.0400$ & $0.00353$ &$2.80\hskip -1 mm\times\hskip -1 mm 10^{-5}$&$66.2^\circ$\\
				$1\,\sigma$	&$\pm 0.10$ &$\pm 1.02$ & $\pm 0.12$& $\pm 1.68$ &$ \pm 0.0007$ &
				$ \pm 0.0008$ & $ \pm 0.00013$ &$^{+0.14}_{-0.12}\hskip -1 mm\times \hskip -1 mm 10^{-5}$&
				$^{+ 3.4^\circ}_{-3.6^\circ}$\\ \hline  
			\end{tabular}
		\end{center}
		\caption{Results of the fits of the quark mass ratios, 
			CKM mixing angles, $J_{\rm CP}$ and $\delta_{\rm CP}$. 'Exp' denotes the  values of the observables 
			at the GUT scale, including $1\sigma$ error.
		}
		\label{tab:output-(1)-3}
	}
\end{table}
%

{ The constants multiplying the modular forms 
present in $M_D$ and in $M_U$ are of the 
	same order of magnitude except for $|c_U/f_U|$, which is somewhat 
	smaller than $|e_U/f_U|$:
}
\begin{align}
\left|\frac{c_D}{f_D}\right |\simeq 0.28\,,\qquad \left|\frac{e_D}{f_D}\right|\simeq 0.59\,,\qquad
\left|\frac{c_U}{f_U}\right|\simeq 0.02\,,\qquad \left|\frac{e_U}{f_U}\right|\simeq 0.14\,.
\end{align}
%
 In what concerns the other constants in $M_D$ and in $M_U$, 
the ratio of $|d_D/a_D|$ is  somewhat smaller than 
$|b_D/a_D|$,
the other ratios $|b_U/a_U|$ and $|d_U/a_U|$
being  of the same order of magnitude:
\begin{align}
\left|\frac{b_D}{a_D}\right |\simeq 0.47\,,\qquad \left|\frac{d_D}{a_D}\right|\simeq 0.058\,,\qquad
\left|\frac{b_U}{a_U}\right|\simeq 1.59\,,\qquad \left|\frac{d_U}{a_U}\right|\simeq 2.50\,.
\end{align}
%
%
\subsection{Fitting  model (2)}
\label{sec:fits-2}

We present three examples of model (2) corresponding to the 
$\tau$ being relatively close to the fixed points $\tau=i$, $\omega$ and
$\infty$, respectively.
%
\subsubsection{$\tau$ close to $i$}
The first example is  close to $\tau=i$. 
For the numerical values of parameters resulting from the fit,
we obtain in this case:
\begin{align}
&{\frac{b_D}{a_D}=-0.0541},\quad {\frac{c_D}{a_D}=0.0212},\quad 
{\frac{d_D}{a_D}=0.871},\quad {\frac{e_D}{a_D}=0.0114},\quad 
{\frac{f_D}{a_D}=0.0157},\nonumber\\
&{\frac{b_U}{a_U}=5.58\times 10^{-3}},\quad 
{\frac{c_U}{a_U}=5.76\times 10^{-2}},\quad 
{\frac{d_U}{a_U}=1.63},\quad {\frac{e_U}{a_U}=5.10\times 10^{-4}},\quad 
{\frac{f_U}{a_U}=1.11\times 10^{-1}},\nonumber\\
&g_D=1.363,\quad g_U=1.151,\quad \tau=0.1706+i\, 1.1417,
 \quad N\sigma=0.82.
\label{}
\end{align}
%
The results of the fit of the observables are shown in  
Table \ref{tab:output-(2)-1}.
\begin{table}[H]
	\small{
		\begin{center}
			\renewcommand{\arraystretch}{1.1}
			\begin{tabular}{|c|c|c|c|c|c|c|c|c|c|} \hline
				\rule[14pt]{0pt}{3pt}  
				& $\frac{m_s}{m_b}\hskip -1 mm\times\hskip -1 mm 10^2$ 
				& $\frac{m_d}{m_b}\hskip -1 mm\times\hskip -1 mm 10^4$& $\frac{m_c}{m_t}\hskip -1 mm\times\hskip -1 mm 10^3$&$\frac{m_u}{m_t}\hskip -1 mm\times\hskip -1 mm 10^6$&
				$|V_{us}|$ &$|V_{cb}|$ &$|V_{ub}|$&$|J_{\rm CP}|$& $\delta_{\rm CP}$
				\\
				\hline
				\rule[14pt]{0pt}{3pt}  
				Fit &$1.87$ & $ 9.10$
				& $2.80$ & $6.47$&
				$0.2250$ & $0.0400$ & $0.00360$ &
				$2.86\hskip -1 mm\times\hskip -1 mm 10^{-5}$&$65.6^\circ$
				\\ \hline
				\rule[14pt]{0pt}{3pt}
				Exp	 &$1.82$ & $9.21$ 
				& $2.80$& $ 5.39$ &
				$0.2250$ & $0.0400$ & $0.00353$ &$2.80\hskip -1 mm\times\hskip -1 mm 10^{-5}$&$66.2^\circ$\\
				$1\,\sigma$	&$\pm 0.10$ &$\pm 1.02$ & $\pm 0.12$& $\pm 1.68$ &$ \pm 0.0007$ &
				$ \pm 0.0008$ & $ \pm 0.00013$ &$^{+0.14}_{-0.12}\hskip -1 mm\times \hskip -1 mm 10^{-5}$&
				$^{+ 3.4^\circ}_{-3.6^\circ}$\\ \hline  
			\end{tabular}
		\end{center}
		\caption{Results of the fits of the quark mass ratios, 
			CKM mixing angles, $J_{\rm CP}$ and $\delta_{\rm CP}$. 
			'Exp' denotes the  values of the observables 
			at the GUT scale, including $1\sigma$ error.
		}
		\label{tab:output-(2)-1}
	}
\end{table}
%
The constants multiplying the modular forms are of the 
same order of magnitude except for $e_U$, which is significantly 
smaller:
\begin{align}
\left|\frac{c_D}{f_D}\right |\simeq 1.35\,,\qquad \left|\frac{e_D}{f_D}\right|\simeq 0.73\,,\qquad
\left|\frac{c_U}{f_U}\right|\simeq 0.52\,,\qquad \left|\frac{e_U}{f_U}\right|\simeq 0.0046\,.
\end{align}
%
The other constants are hierarchical:
\begin{align}
\left|\frac{b_D}{a_D}\right |\simeq 0.054\,,\qquad \left|\frac{d_D}{a_D}\right|\simeq 0.87\,,\qquad
\left|\frac{b_U}{a_U}\right|\simeq 0.0056\,,\qquad \left|\frac{d_U}{a_U}\right|\simeq 1.63\,.
\end{align}
%

%
\subsubsection{$\tau$ close to $\omega$}
%
%

We present  the second example, in which the VEV of the modulus $\tau$ 
is close to the fixed point $\omega$.
The numerical values of parameters obtained in the fit are:
\begin{align}
&{\frac{b_D}{a_D}=0.0505},\quad {\frac{c_D}{a_D}=-0.278},\quad 
{\frac{d_D}{a_D}=0.839},\quad {\frac{e_D}{a_D}=0.0317},\quad 
{\frac{f_D}{a_D}=-0.00963},\nonumber\\
&{\frac{b_U}{a_U}=4.75\times 10^{-3}},\quad 
{\frac{c_U}{a_U}=0.889},\quad 
{\frac{d_U}{a_U}=1.72},\quad {\frac{e_U}{a_U}=-1.89\times 10^{-3}},\quad 
{\frac{f_U}{a_U}=-6.34\times 10^{-2}},\nonumber\\
& g_D=1.64,\ \, g_U=-0.840, \ \, \tau=-0.4474+i\,0.9357,\ \, \tau-\omega=0.0526+ i\,0.0697,  \ \, N\sigma=0.96.
\label{}
\end{align}
%
The results of the fit of the observables are shown in  
Table \ref{tab:output-(2)-2}.
\begin{table}[H]
	\small{
		\begin{center}
			\renewcommand{\arraystretch}{1.1}
			\begin{tabular}{|c|c|c|c|c|c|c|c|c|c|} \hline
				\rule[14pt]{0pt}{3pt}  
				& $\frac{m_s}{m_b}\hskip -1 mm\times\hskip -1 mm 10^2$ 
				& $\frac{m_d}{m_b}\hskip -1 mm\times\hskip -1 mm 10^4$& $\frac{m_c}{m_t}\hskip -1 mm\times\hskip -1 mm 10^3$&$\frac{m_u}{m_t}\hskip -1 mm\times\hskip -1 mm 10^6$&
				$|V_{us}|$ &$|V_{cb}|$ &$|V_{ub}|$&$|J_{\rm CP}|$& $\delta_{\rm CP}$
				\\
				\hline
				\rule[14pt]{0pt}{3pt}  
				Fit &$1.89$ & $ 9.27$
				& $2.86$ & $4.48$&
				$0.2252$ & $0.0400$ & $0.00349$ &
				$2.76\hskip -1 mm\times\hskip -1 mm 10^{-5}$&$64.7^\circ$
				\\ \hline
				\rule[14pt]{0pt}{3pt}
				Exp	 &$1.82$ & $9.21$ 
				& $2.80$& $ 5.39$ &
				$0.2250$ & $0.0400$ & $0.00353$ &$2.80\hskip -1 mm\times\hskip -1 mm 10^{-5}$&$66.2^\circ$\\
				$1\,\sigma$	&$\pm 0.10$ &$\pm 1.02$ & $\pm 0.12$& $\pm 1.68$ &$ \pm 0.0007$ &
				$ \pm 0.0008$ & $ \pm 0.00013$ &$^{+0.14}_{-0.12}\hskip -1 mm\times \hskip -1 mm 10^{-5}$&
				$^{+ 3.4^\circ}_{-3.6^\circ}$\\ \hline  
			\end{tabular}
		\end{center}
		\caption{Results of the fits of the quark mass ratios, 
			CKM mixing angles, $J_{\rm CP}$ and $\delta_{\rm CP}$. 
			'Exp' denotes the  values of the observables 
			at the GUT scale, including $1\sigma$ error.
		}
		\label{tab:output-(2)-2}
	}
\end{table}
%
The constants multiplying the modular forms are of the 
same order of magnitude except for $e_U$, which is significantly 
smaller also in this case:
\begin{align}
\left|\frac{c_D}{f_D}\right |\simeq 28\,,\qquad \left|\frac{e_D}{f_D}\right|\simeq 3.3\,,\qquad
\left|\frac{c_U}{f_U}\right|\simeq 14\,,\qquad \left|\frac{e_U}{f_U}\right|\simeq 0.030\,.
\end{align}
%
The other constants are hierarchical:
\begin{align}
\left|\frac{b_D}{a_D}\right |\simeq 0.051\,,\qquad \left|\frac{d_D}{a_D}\right|\simeq 0.84\,,\qquad
\left|\frac{b_U}{a_U}\right|\simeq 0.0048\,,\qquad \left|\frac{d_U}{a_U}\right|\simeq 1.72\,.
\end{align}
%
\subsubsection{$\tau$ "close" to $i\infty$}
%

We present the third  example at the VEV of  $\tau$ 
having a relatively large imaginary part ``close'' to  $i\infty$.
The numerical values of parameters obtained in the fit are: 
\begin{align}
&{\frac{b_D}{a_D}=1.07\times 10^{-2}},\ \ {\frac{c_D}{a_D}=-4.96\times 10^{-3}},\ \
{\frac{d_D}{a_D}=-0.166},\ \ {\frac{e_D}{a_D}=2.35\times 10^{-3}},\ \ 
{\frac{f_D}{a_D}=5.38\times 10^{-3}},\nonumber\\
&{\frac{b_U}{a_U}=1.58\times 10^{-4}},\ \
{\frac{c_U}{a_U}=2.99\times 10^{-3}},\ \ 
{\frac{d_U}{a_U}=0.465},\ \ {\frac{e_U}{a_U}=-6.98\times 10^{-6}},\ \ 
{\frac{f_U}{a_U}=-4.11\times 10^{-2}},\nonumber\\
& g_D=0.124,\quad g_U=-0.0709, \quad \tau=0.3102+i\,1.5672,\quad N\sigma=0.774.
\label{}
\end{align}
%
The results of the fit of the observables are shown in  
Table \ref{tab:output-(2)-3}.
\begin{table}[H]
	\small{
		\begin{center}
			\renewcommand{\arraystretch}{1.1}
			\begin{tabular}{|c|c|c|c|c|c|c|c|c|c|} \hline
				\rule[14pt]{0pt}{3pt}  
				& $\frac{m_s}{m_b}\hskip -1 mm\times\hskip -1 mm 10^2$ 
				& $\frac{m_d}{m_b}\hskip -1 mm\times\hskip -1 mm 10^4$& $\frac{m_c}{m_t}\hskip -1 mm\times\hskip -1 mm 10^3$&$\frac{m_u}{m_t}\hskip -1 mm\times\hskip -1 mm 10^6$&
				$|V_{us}|$ &$|V_{cb}|$ &$|V_{ub}|$&$|J_{\rm CP}|$& $\delta_{\rm CP}$
				\\
				\hline
				\rule[14pt]{0pt}{3pt}  
				Fit &$1.87$ & $ 9.79$
				& $2.79$ & $5.96$&
				$0.2250$ & $0.0402$ & $0.00352$ &
				$2.82\hskip -1 mm\times\hskip -1 mm 10^{-5}$&$65.3^\circ$
				\\ \hline
				\rule[14pt]{0pt}{3pt}
				Exp	 &$1.82$ & $9.21$ 
				& $2.80$& $ 5.39$ &
				$0.2250$ & $0.0400$ & $0.00353$ &$2.80\hskip -1 mm\times\hskip -1 mm 10^{-5}$&$66.2^\circ$\\
				$1\,\sigma$	&$\pm 0.10$ &$\pm 1.02$ & $\pm 0.12$& $\pm 1.68$ &$ \pm 0.0007$ &
				$ \pm 0.0008$ & $ \pm 0.00013$ &$^{+0.14}_{-0.12}\hskip -1 mm\times \hskip -1 mm 10^{-5}$&
				$^{+ 3.4^\circ}_{-3.6^\circ}$\\ \hline  
			\end{tabular}
		\end{center}
		\caption{Results of the fits of the quark mass ratios, 
			CKM mixing angles, $J_{\rm CP}$ and $\delta_{\rm CP}$. 
			'Exp' denotes the  values of the observables 
			at the GUT scale, including $1\sigma$ error.
		}
		\label{tab:output-(2)-3}
	}
\end{table}
%
The constants multiplying the modular forms are of the 
same order of magnitude for the down-type quarks,
but hierarchical for up-type quarks:
\begin{align}
\left|\frac{c_D}{f_D}\right |\simeq 0.92\,,\qquad \left|\frac{e_D}{f_D}\right|\simeq 0.44\,,\qquad
\left|\frac{c_U}{f_U}\right|\simeq 0.073\,,\qquad \left|\frac{e_U}{f_U}\right|\simeq 0.00017\,.
\end{align}
%
%
The other constants are hierarchical:
\begin{align}
\left|\frac{b_D}{a_D}\right |\simeq 0.011\,,\qquad \left|\frac{d_D}{a_D}\right|\simeq 0.17\,,\qquad
\left|\frac{b_U}{a_U}\right|\simeq 0.00016\,,\qquad \left|\frac{d_U}{a_U}\right|\simeq 0.47\,.
\end{align}
%

%
\subsection{Fitting  model (3)}
\label{sec:fits-3}
%
%
We present three examples of model (3) corresponding to the 
$\tau$ being relatively close to the fixed points $\tau=i$, $\omega$ and
 $\infty$, respectively.
%
\subsubsection{$\tau$ close to $i$}
%
%
The first one is an  example of model (3), where
$\tau$ is rather close to $i$. 
For the numerical values of parameters resulting from the fit,
we obtain in this case:
\begin{align}
&{\frac{a_D}{d_D}=-9.38\times 10^{-4}},\quad {\frac{b_D}{d_D}=-1.83\times 10^{-2}},\quad 
{\frac{c_D}{d_D}=2.66\times 10^{-5}},\quad {\frac{e_D}{d_D}=7.65\times 10^{-3}},\nonumber\\
&{\frac{f_D}{d_D}=3.58\times 10^{-6}},\quad
{\frac{a_U}{d_U}=5.39\times 10^{-6}},\quad {\frac{b_U}{d_U}=2.80\times 10^{-3}},\quad 
{\frac{c_U}{d_U}=0},\quad
{\frac{e_U}{d_U}=0},\quad 
{\frac{f_U}{d_U}=0},\nonumber\\
& g_D=1.199,\quad \tau=0.2505+i\, 1.1356 ,
\quad N\sigma=0.15. 
\label{}
\end{align}
%
Both the constants multiplying the modular forms present in $M_D$ 
and the other constants in $M_D$ and $M_U$ have very 
different values with the their relevant ratios exhibiting strong 
hierarchies.
The results of the fit of the observables are shown in  
Table \ref{tab:output-(3)-1}.
\begin{table}[H]
	\small{
		\begin{center}
			\renewcommand{\arraystretch}{1.1}
			\begin{tabular}{|c|c|c|c|c|c|c|c|c|c|} \hline
				\rule[14pt]{0pt}{3pt}  
				& $\frac{m_s}{m_b}\hskip -1 mm\times\hskip -1 mm 10^2$ 
				& $\frac{m_d}{m_b}\hskip -1 mm\times\hskip -1 mm 10^4$& $\frac{m_c}{m_t}\hskip -1 mm\times\hskip -1 mm 10^3$&$\frac{m_u}{m_t}\hskip -1 mm\times\hskip -1 mm 10^6$&
				$|V_{us}|$ &$|V_{cb}|$ &$|V_{ub}|$&$|J_{\rm CP}|$& $\delta_{\rm CP}$
				\\
				\hline
				\rule[14pt]{0pt}{3pt}  
				Fit &$1.87$ & $ 9.13$
				& $2.80$ & $5.39$&
				$0.2250$ & $0.0400$ & $0.00353$ &
				$2.84\hskip -1 mm\times\hskip -1 mm 10^{-5}$&$66.5^\circ$
				\\ \hline
				\rule[14pt]{0pt}{3pt}
				Exp	 &$1.82$ & $9.21$ 
				& $2.80$& $ 5.39$ &
				$0.2250$ & $0.0400$ & $0.00353$ &$2.80\hskip -1 mm\times\hskip -1 mm 10^{-5}$&$66.2^\circ$\\
				$1\,\sigma$	&$\pm 0.10$ &$\pm 1.02$ & $\pm 0.12$& $\pm 1.68$ &$ \pm 0.0007$ &
				$ \pm 0.0008$ & $ \pm 0.00013$ &$^{+0.14}_{-0.12}\hskip -1 mm\times \hskip -1 mm 10^{-5}$&
				$^{+ 3.4^\circ}_{-3.6^\circ}$\\ \hline  
			\end{tabular}
		\end{center}
		\caption{Results of the fits of the quark mass ratios, 
			CKM mixing angles, $J_{\rm CP}$ and $\delta_{\rm CP}$. 
			'Exp' denotes the  values of the observables 
			at the GUT scale, including $1\sigma$ error.
		}
		\label{tab:output-(3)-1}
	}
\end{table}
%

%
\subsubsection{$\tau$ close to $\omega$ }
%
%
We present the second example, in which the VEV of the modulus $\tau$ 
is close to the fixed point $\omega$.
The numerical values of parameters obtained in the fit are: 
\begin{align}
&{\frac{a_D}{d_D}=9.68\times 10^{-4}},\quad {\frac{b_D}{d_D}=1.82\times 10^{-2}},\quad 
{\frac{c_D}{d_D}=4.00\times 10^{-4}},\quad {\frac{e_D}{d_D}=-3.04\times 10^{-2}},\nonumber \\
&{\frac{f_D}{d_D}=-2.65\times 10^{-6}},\quad
{\frac{a_U}{d_U}=5.39\times 10^{-6}},\quad {\frac{b_U}{d_U}=2.80\times 10^{-3}},\quad 
{\frac{c_U}{d_U}=0},\quad
{\frac{e_U}{d_U}=0},\quad 
{\frac{f_U}{d_U}=0},\nonumber\\
&g_D=-0.869, \quad \tau=-0.4706+i\, 0.9359 , \quad \tau-\omega=0.0295+i\, 0.0699,
\quad N\sigma=0.24.
\label{}
\end{align}
%
As in the previous examples, there is a strong hierarchy 
among both the constants multiplying the modular forms present in $M_D$ 
and the other constants in $M_D$ and $M_U$.

The results of the fit of the observables are shown in  
Table \ref{tab:output-(3)-2}.
\begin{table}[H]
	\small{
		\begin{center}
			\renewcommand{\arraystretch}{1.1}
			\begin{tabular}{|c|c|c|c|c|c|c|c|c|c|} \hline
				\rule[14pt]{0pt}{3pt}  
				& $\frac{m_s}{m_b}\hskip -1 mm\times\hskip -1 mm 10^2$ 
				& $\frac{m_d}{m_b}\hskip -1 mm\times\hskip -1 mm 10^4$& $\frac{m_c}{m_t}\hskip -1 mm\times\hskip -1 mm 10^3$&$\frac{m_u}{m_t}\hskip -1 mm\times\hskip -1 mm 10^6$&
				$|V_{us}|$ &$|V_{cb}|$ &$|V_{ub}|$&$|J_{\rm CP}|$& $\delta_{\rm CP}$
				\\
				\hline
				\rule[14pt]{0pt}{3pt}  
				Fit &$1.86$ & $ 9.43$
				& $2.80$ & $5.39$&
				$0.2250$ & $0.0400$ & $0.00354$ &
				$2.84\hskip -1 mm\times\hskip -1 mm 10^{-5}$&$66.4^\circ$
				\\ \hline
				\rule[14pt]{0pt}{3pt}
				Exp	 &$1.82$ & $9.21$ 
				& $2.80$& $ 5.39$ &
				$0.2250$ & $0.0400$ & $0.00353$ &$2.80\hskip -1 mm\times\hskip -1 mm 10^{-5}$&$66.2^\circ$\\
				$1\,\sigma$	&$\pm 0.10$ &$\pm 1.02$ & $\pm 0.12$& $\pm 1.68$ &$ \pm 0.0007$ &
				$ \pm 0.0008$ & $ \pm 0.00013$ &$^{+0.14}_{-0.12}\hskip -1 mm\times \hskip -1 mm 10^{-5}$&
				$^{+ 3.4^\circ}_{-3.6^\circ}$\\ \hline  
			\end{tabular}
		\end{center}
		\caption{Results of the fits of the quark mass ratios, 
			CKM mixing angles, $J_{\rm CP}$ and $\delta_{\rm CP}$. 
			'Exp' denotes the  values of the observables 
			at the GUT scale, including $1\sigma$ error.
		}
		\label{tab:output-(3)-2}
	}
\end{table}
%
%
\subsubsection{$\tau$ ''close'' to $i\infty$}
%
%
Finally, we present an  example at the VEV of  $\tau$ 
having a relatively large imaginary part ``close'' to  $i\infty$.
The numerical values of parameters obtained in the fit are: 
\begin{align}
&{\frac{a_D}{d_D}=9.45\times 10^{-4}},\quad 
{\frac{b_D}{d_D}=-1.83\times 10^{-2}},\quad 
{\frac{c_D}{d_D}=-8.22\times 10^{-6}},\quad {\frac{e_D}{d_D}=-3.54\times 10^{-3}},\nonumber\\
&{\frac{f_D}{d_D}=-1.28\times 10^{-5}},\quad
{\frac{a_U}{d_U}=5.39\times 10^{-6}},\quad {\frac{b_U}{d_U}=2.80\times 10^{-3}},\quad 
{\frac{c_U}{d_U}=0},\quad
{\frac{e_U}{d_U}=0},\quad 
{\frac{f_U}{d_U}=0},\nonumber\\
&g_D=0.00925,\quad \tau=-0.288076+i\, 1.68188,
\quad N\sigma=0.097. 
\label{}
\end{align}
%
In this example, there is also  a strong hierarchy 
among both the constants multiplying the modular forms present in $M_D$ 
and the other constants in $M_D$ and $M_U$.

The results of the fit of the observables are shown in  
Table \ref{tab:output-(3)-3}.
\begin{table}[H]
	\small{
		\begin{center}
			\renewcommand{\arraystretch}{1.1}
			\begin{tabular}{|c|c|c|c|c|c|c|c|c|c|} \hline
				\rule[14pt]{0pt}{3pt}  
				& $\frac{m_s}{m_b}\hskip -1 mm\times\hskip -1 mm 10^2$ 
				& $\frac{m_d}{m_b}\hskip -1 mm\times\hskip -1 mm 10^4$& $\frac{m_c}{m_t}\hskip -1 mm\times\hskip -1 mm 10^3$&$\frac{m_u}{m_t}\hskip -1 mm\times\hskip -1 mm 10^6$&
				$|V_{us}|$ &$|V_{cb}|$ &$|V_{ub}|$&$|J_{\rm CP}|$& $\delta_{\rm CP}$
				\\
				\hline
				\rule[14pt]{0pt}{3pt}  
				Fit &$1.87$ & $ 9.24$
				& $2.80$ & $5.39$&
				$0.2250$ & $0.0400$ & $0.00352$ &
				$2.83\hskip -1 mm\times\hskip -1 mm 10^{-5}$&$66.4^\circ$
				\\ \hline
				\rule[14pt]{0pt}{3pt}
				Exp	 &$1.82$ & $9.21$ 
				& $2.80$& $ 5.39$ &
				$0.2250$ & $0.0400$ & $0.00353$ &$2.80\hskip -1 mm\times\hskip -1 mm 10^{-5}$&$66.2^\circ$\\
				$1\,\sigma$	&$\pm 0.10$ &$\pm 1.02$ & $\pm 0.12$& $\pm 1.68$ &$ \pm 0.0007$ &
				$ \pm 0.0008$ & $ \pm 0.00013$ &$^{+0.14}_{-0.12}\hskip -1 mm\times \hskip -1 mm 10^{-5}$&
				$^{+ 3.4^\circ}_{-3.6^\circ}$\\ \hline  
			\end{tabular}
		\end{center}
		\caption{Results of the fits of the quark mass ratios, 
			CKM mixing angles, $J_{\rm CP}$ and $\delta_{\rm CP}$. 
			'Exp' denotes the  values of the observables 
			at the GUT scale, including $1\sigma$ error.
		}
		\label{tab:output-(3)-3}
	}
\end{table}
%
\section{A texture realizing modulus stabilisation}
\subsection{Modulus stabilisation}
We have performed statistical analyses of the three models 
and have shown that they are phenomenologically viable. 
We have shown that model (1) (Eq.\,(\ref{model-1})),
model  (2) (Eq.\,(\ref{model-2})
and model (3) (Eq.\,(\ref{model-3-PH}) and the related discussion) 
can describe well the quark data for values of 
$\tau_{\rm VEV}$ close to $i$, 
$\tau_{\rm VEV}$ close to $\omega = -\,0.5 + i\,\sqrt{3}/2$ 
(the left cusp) and $\tau_{\rm VEV}$ ``close'' to $i\infty$.
The values of the VEV of the modulus $\tau$,
$\tau_{\rm VEV} = i$, $\tau_{\rm VEV} = \omega$
and $\tau_{\rm VEV} = i\infty$, as is well known, 
are the only fixed points of the modular group 
in its fundamental domain. 
Values of $\tau_{\rm VEV}$ close to three fixed points 
have been found in studies of the modulus stabilisation 
(see, e.g., \cite{Kobayashi:2019xvz,Abe:2020vmv,Ishiguro:2020tmo,Novichkov:2022wvg,Ishiguro:2022pde,Knapp-Perez:2023nty,King:2023snq,Kobayashi:2023spx,Higaki:2024jdk}), in which $\tau_{\rm VEV}$ is obtained   
by analysing the absolute minima of 
relatively simple (supergravity-motivated) 
modular- and CP-invariant potentials for the modulus $\tau$. 
 In this section, we focus on  the solution,
 which give the absolute minima of the supergravity-motivated
 modular- and CP-invariant potentials for the modulus $\tau$,
 obtained in Ref.\cite{Novichkov:2022wvg}.
 \begin{table}[h]
 	\begin{center}
 		\renewcommand{\arraystretch}{1.1}
 		\begin{tabular}{|c|c|} \hline
 		&  $\tau_{\rm VEV}$ \\ \hline
 	$n=0$, \	$m=1$	& $-0.484+0.884 \,i$ \\ 
 		$n=0$,	\	$m=2$	& $-0.492+0.875\,i$ \\ 
 			$n=0$,	\	$m=3$	& $-0.495+0.872 \,i$ \\ \hline
 		\end{tabular}
 	\end{center}
 	\caption{Values of the modulus  $\tau_{\rm VEV}$ at the global minima of the the supergravity-motivated modular- and CP-invariant potentials in Ref.\cite{Novichkov:2022wvg}.
 	Here, $(m,n)$ denote the power indices in the modular-invariant function. }
 	\label{tab:minima}
 \end{table}
 The values of $\tau_{\rm VEV}$ are very close to the  fixed point $\omega$ for the non-negative integer $(m,n)$,
  which denote the power indices in the modular-invariant function,
 as seen in Table \ref{tab:minima}.

 In the next subsection, we present a model, which is consistent with $\tau_{\rm VEV}$ in 
 Table \ref{tab:minima}.

\subsection{An alternative texture zero }
In the standpoint of "Occam's Razor approach" of the quark mass matrix
(minimum number of parameters)
\cite{Tanimoto:2016rqy}, we consider the following texture zeros for mass matrices of the down-type quarks with the diagonal  up-type quark mass matrix.
\begin{align}
&  M_D =v_D
\begin{pmatrix}
0  & 0 & a_D  \\
0&b_D & c_D \\
d_D& e_D& f_D  \, e^{-i \phi_D}\end{pmatrix}_{RL},
\qquad
M_U =v_U
\begin{pmatrix}
a_U  & 0 &0 \\
0&b_U & 0 \\
0& 0& d_U  \\
\end{pmatrix}_{RL}, 
\label{texture-0}
\end{align}
where $a_D,\, b_D,\, c_D,\, d_D,\, e_D,\,f_D,\,a_U,\, b_U,\, c_U,\, d_U,\, e_U$ and $f_U$ are real, and $ \phi_D$ is the CP phase.
For down-type quark sector, we obtain those numerical values
by inputting observed ones in Appendix \ref{Appen-inputs} as follows:

\begin{align}
&\frac{b_D}{a_D}=0.2646,\quad \frac{c_D}{a_D}=7.0815,\quad 
\frac{d_D}{a_D}=0.05156,\quad \frac{e_D}{a_D}=0.3092,\quad 
\frac{f_D}{a_D}=5.8735,\nonumber\\
&\varphi_D=-43.55^\circ\,,\qquad N\sigma=0.0433\,.
\end{align}

\begin{table}[H]
	\small{
		\begin{center}
			\renewcommand{\arraystretch}{1.1}
			\begin{tabular}{|c|c|c|c|c|c|c|c|c|c|} \hline
				\rule[14pt]{0pt}{3pt}  
				& $\frac{m_s}{m_b}\hskip -1 mm\times\hskip -1 mm 10^2$ 
				& $\frac{m_d}{m_b}\hskip -1 mm\times\hskip -1 mm 10^4$& $\frac{m_c}{m_t}\hskip -1 mm\times\hskip -1 mm 10^3$&$\frac{m_u}{m_t}\hskip -1 mm\times\hskip -1 mm 10^6$&
				$|V_{us}|$ &$|V_{cb}|$ &$|V_{ub}|$&$|J_{\rm CP}|$& $\delta_{\rm CP}$
				\\
				\hline
				\rule[14pt]{0pt}{3pt}  
				Fit &$1.87$ & $ 9.19$
				& *  & * &
				$0.2250$ & $0.0400$ & $0.00353$ &
				$2.83\hskip -1 mm\times\hskip -1 mm 10^{-5}$&$66.2^\circ$
				\\ \hline
				\rule[14pt]{0pt}{3pt}
				Exp	 &$1.82$ & $9.21$ 
				& $2.80$& $ 5.39$ &
				$0.2250$ & $0.0400$ & $0.00353$ &$2.8\hskip -1 mm\times\hskip -1 mm 10^{-5}$&$66.2^\circ$\\
				$1\,\sigma$	&$\pm 0.10$ &$\pm 1.02$ & $\pm 0.12$& $\pm 1.68$ &$ \pm 0.0007$ &
				$ \pm 0.0008$ & $ \pm 0.00013$ &$^{+0.14}_{-0.12}\hskip -1 mm\times \hskip -1 mm 10^{-5}$&
				$^{+ 3.4^\circ}_{-3.6^\circ}$\\ \hline  
			\end{tabular}
		\end{center}
		\caption{Results of the fits of the quark mass ratios, 
			CKM mixing angles, $J_{\rm CP}$ and $\delta_{\rm CP}$. 'Exp' denotes the  values of the observables 
			at the GUT scale, including $1\sigma$ error.
		}
		\label{tab:output0}
	}
\end{table}

It is emphasized that 
the numerical values of the parameter set are unique in these mass matrices
because  the number of parameters is ten while input data are also ten.

%

\subsection{ Variant of model (1)}
We consider the variant model of the model (1) in Table \ref{tab:model-1}.
We show the assignment of the representations and weights for the relevant quarks
in Table \ref{tab:variant},
where the assignments of $(u^c,c^c,t^c)$ are changed:
\begin{table}[H]
	\begin{center}
		\renewcommand{\arraystretch}{1.1}
		\begin{tabular}{|c|c|c|c|c|} \hline
			& $(d,u)_L,(s,c)_L,(b,t)_L$ & $(d^c,s^c,b^c),\,(u^c,c^c,t^c)$ &  $H_U$ & $H_D$ \\ \hline
			$SU(2)$ & 2 & 1 \quad\qquad 1  & 2 & 2 \\
			$A_4$ & $(1,\,1\,,1'')$ &  $(1',\,1\,,1)$\ \  $(1,\,1\,,1')$ & $1$ & $1$ \\
			$k$ &$ (-8,-4,\ 8)$  &$(-8,4,\ 8)$ $(8,4,\ -8)$ & 0 & 0 \\ \hline
		\end{tabular}
	\end{center}
	\caption{Assignments of $A_4$ representations and weights.  }
	\label{tab:variant}
\end{table}
Since the representations and the weights of the right-handed up-type quarks are different from the  down-type ones,
the mass matrices are different each other.
We have real ${\rm det }\,[M_D]=-a_D b_D d_D$ and 
${\rm det }\, [M_U]=a_U b_U d_U$.
The mass matrices of the down- and up-type quarks are given as:
\begin{align}
M_D =v_D
\begin{pmatrix}
0  & 0 & {a_D} \\
0&{b_D}& {c_Q}  (2 {\rm Im} \tau)^{6} Y_{\bf 1'}^{(12)}\\
d_D&{e_D}  (2 {\rm Im} \tau)^{2} Y_{\bf 1}^{(4)}&
{f_D} (2 {\rm Im} \tau)^{8}(g_D Y_{\bf 1'_{\rm A}}^{(16)}+
Y_{\bf 1'_{\rm B}}^{(16)}) \end{pmatrix}_{RL} \,,
\nonumber \\
M_U =v_U
\begin{pmatrix}
a_U&{e_U}  (2 {\rm Im} \tau)^{2} Y_{\bf 1}^{(4)}&
{f_U} (2 {\rm Im} \tau)^{8}(g_U Y_{\bf 1'_{\rm A}}^{(16)}+
Y_{\bf 1'_{\rm B}}^{(16)}) \\
0&{b_U}& {c_U}  (2 {\rm Im} \tau)^{6} Y_{\bf 1'}^{(12)}\\
0  & 0 & {d_U}
\end{pmatrix}_{RL} \,,
\label{variant}
\end{align}
In order to reproduce the texture zeros in \eqref{texture-0},
we choose $c_U=e_U=f_U=0$ or $c_U,e_U,f_U<<a_U,b_U,d_U$ by hand.
%

\subsection{Numerical result of the variant model}
In the quark mass matrices of Eq.\eqref{variant}, the allowed $\tau$ region is obtained 
by fitting the phase of the texture in \eqref{texture-0} with 
error bar $1\%$ as discussed in section \ref{A4}.
We show the full allowed region in Fig.\ref{variant-Full} and 
the restricted region close to the fixed point $\tau=\omega$ in
Fig.\ref{variant-Micro}.

\begin{figure}[H]
	\begin{tabular}{ccc}
		\begin{minipage}{0.47\hsize}
			\includegraphics[width=\linewidth]{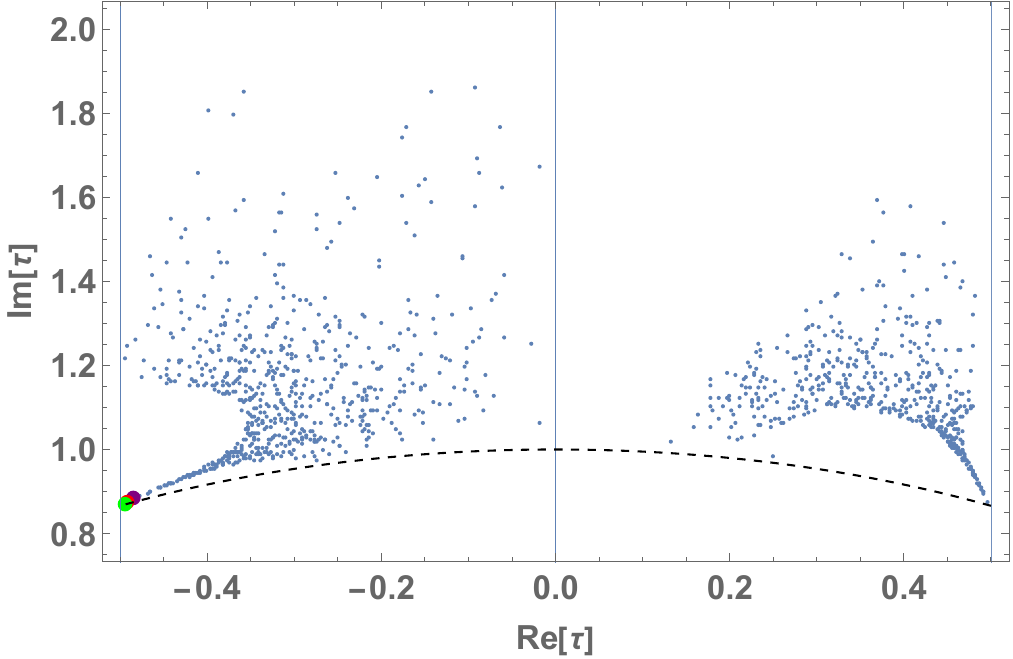}
			\caption{Allowed region in the fundamental domain.
				Purple, red and green points close to $\omega$ denote
				the $\tau$ in the cases of 
				$m=1,2,3$ with $n=0$, respectively.
			The dotted curves denotes the boundary of $|\tau|=1$.	}
			\label{variant-Full}
		\end{minipage}
		\hskip 0.7 cm
		\begin{minipage}{0.47\hsize}
		\includegraphics[width=\linewidth]{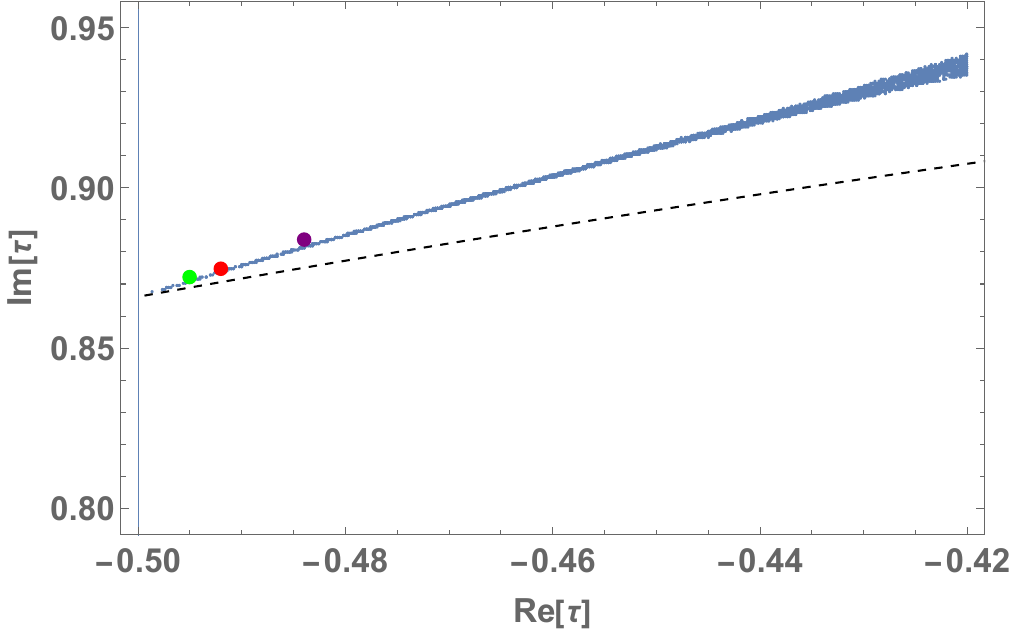}
			\caption{Allowed region close to $\omega$.
				Purple, red and green points denote
				the $\tau$ in the cases of 
				$m=1,2,3$ with $n=0$, respectively.
			The dotted curves denotes the boundary of $|\tau|=1$.	}
			\label{variant-Micro}
		\end{minipage}
	\end{tabular}
\end{figure}
It is  remarked that the three points in
Table \ref{tab:minima} are almost on the  line, which is very narrow allowed region.
These three points  are never inside the allowed region in the previous models  as seen in Figs.\ref{tau-(1)},
\ref{tau-Eisen}, \ref{tau-(2)}, \ref{tau-(3)}.

Indeed, we obtain a successful  example fitting quark masses and CKM parameters
by fixing $\tau=-0.492+0.875 \,i$,
which corresponds to  $(m.n)=(2,0)$ in Table \ref{tab:minima} as follows:
\begin{align}
&\frac{b_D}{a_D}=0.2717,\quad \frac{c_D}{a_D}=-78.09,\quad 
\frac{d_D}{a_D}=0.05323,\quad \frac{e_D}{a_D}=-1.431,\quad 
\frac{f_D}{a_D}=0.01308,\nonumber\\
&g_D=-1.490\,,\quad \tau=-0.492+0.875 \,i\,,\quad N\sigma=0.861\,,
\end{align}
where the up-type quark mass matrix is diagonal.
The output is given 
in Table \ref{tab:add2}.
\begin{table}[H]
	\small{
		\begin{center}
			\renewcommand{\arraystretch}{1.1}
			\begin{tabular}{|c|c|c|c|c|c|c|c|c|c|} \hline
				\rule[14pt]{0pt}{3pt}  
				& $\frac{m_s}{m_b}\hskip -1 mm\times\hskip -1 mm 10^2$ 
				& $\frac{m_d}{m_b}\hskip -1 mm\times\hskip -1 mm 10^4$& $\frac{m_c}{m_t}\hskip -1 mm\times\hskip -1 mm 10^3$&$\frac{m_u}{m_t}\hskip -1 mm\times\hskip -1 mm 10^6$&
				$|V_{us}|$ &$|V_{cb}|$ &$|V_{ub}|$&$|J_{\rm CP}|$& $\delta_{\rm CP}$
				\\
				\hline
				\rule[14pt]{0pt}{3pt}  
				Fit &$1.94$ & $ 9.11$
				& * & * &
				$0.2250$ & $0.0398$ & $0.00354$ &
				$2.85\hskip -1 mm\times\hskip -1 mm 10^{-5}$&$67.6^\circ$
				\\ \hline
				\rule[14pt]{0pt}{3pt}
				Exp	 &$1.82$ & $9.21$ 
				& $2.80$& $ 5.39$ &
				$0.2250$ & $0.0400$ & $0.00353$ &$2.8\hskip -1 mm\times\hskip -1 mm 10^{-5}$&$66.2^\circ$\\
				$1\,\sigma$	&$\pm 0.10$ &$\pm 1.02$ & $\pm 0.12$& $\pm 1.68$ &$ \pm 0.0007$ &
				$ \pm 0.0008$ & $ \pm 0.00013$ &$^{+0.14}_{-0.12}\hskip -1 mm\times \hskip -1 mm 10^{-5}$&
				$^{+ 3.4^\circ}_{-3.6^\circ}$\\ \hline  
			\end{tabular}
		\end{center}
		\caption{Results of the fits of the quark mass ratios, 
			CKM mixing angles, $J_{\rm CP}$ and $\delta_{\rm CP}$. 'Exp' denotes the  values of the observables 
			at the GUT scale, including $1\sigma$ error.
		}
		\label{tab:add2}
	}
\end{table}

\section{Summary and discussions}
%

 We have discussed the strong CP problem within the modular 
invariance approach to flavour. Working with $A_4$ ($N=3$) modular 
symmetry we have constructed simple models which 
provide solution to the strong CP problem without the need for an axion.
In these models it is assumed that CP is a fundamental symmetry 
of the Lagrangian. As a consequence, the strong CPV  phase 
$\bar{\theta} = 0$. The CP symmetry is broken spontaneously by the VEV of the  modulus $\tau$ {($\tau_{\rm VEV}$)}, so the large CPV phases in the CKM matrix is generated and at the same time $\bar{\theta}$ remains zero or gets a tiny value compatible with the existing stringent experimental limit 
$\bar{\theta} < 10^{-10}$. 

 To be more specific, we have considered three models, i.e., 
three types of down-type and up-type mass matrices $M_D$ and $M_U$, 
which have, as a consequence of the $A_4$ modular symmetry, 
three zero elements, or three texture zeros, each 
(Eqs.\,(\ref{model-1}),  (\ref{model-2}) and (\ref{model-3-PH})). 
The position of the zeros and the requirement of CP invariance ensures that 
${\rm det}M_D$ and ${\rm det}M_U$ are real quantities.
The quark mass matrices $M_D$ and $M_U$ contain three 
$A_4$ modular forms of weights 12 and 16 which are 
${\bf 1'}$ singlets, $Y^{(12)}_{1'}$, $Y^{(16)}_{1'A}$ and  $Y^{(16)}_{1'B}$, 
and one modular form of weight 4 which is ${\bf 1}$ singlet, 
$Y^{(4)}_{1}$. The presence of two modular forms of the same weight 
furnishing the same singlet representation of $A_4$ is required 
in order to describe correctly the observed CP violation 
in the quark sector when the CP symmetry is broken spontaneously 
by $\tau_{\rm VEV}$. In the case of model (3) we have considered 
phenomenologically the presence of three additional zero elements 
in $M_U$ to reduce the number of redundant constant parameters 
in the model. This set-up, in which $M_U$ is a diagonal matrix, 
is not guaranteed by modular invariance without   
additional (symmetry) assumptions.

Our work is an extension of the 
pioneering work of Ref.\cite{Feruglio:2023uof} 
where  one pattern  of  texture zeros is 
considered in the framework of the level $N=1$ modular symmetry.
We have presented phenomenologically viable models with  
finite modular symmetry of level $N=3$ ($A_4$) 
by using three patterns of texture zeros.
We have studied the distributions of the VEV of 
the modulus $\tau$ allowing to reproduce the observed quark masses 
CKM mixing angles and CP violation. 
And we have found examples 
of the models which are viable for values of $\tau_{\rm VEV}$ 
close to the fixed points of the modular group. 

 In particular, we focus  on the VEV of $\tau$,  
	which gives the absolute minima of the supergravity-motivated
	modular- and CP-invariant potentials for the modulus $\tau$,
	so called, modulus stabilisation
	\cite{Novichkov:2022wvg}. We present a successful model, which is  consistent with this results of the  modulus stabilisation.

 Our work together with Ref.\cite{Feruglio:2023uof} 
promotes the modular invariance as a successful approach and 
framework providing solutions not only to the quark and lepton 
flavour problems but also to the strong CP problem.

%
\section*{Acknowledgments}
%
%
The work of S. T. P. was supported in part by the European
Union's Horizon 2020 research and innovation programme under the 
Marie Sk\l{}odowska-Curie grant agreement No.~860881-HIDDeN, by the Italian 
INFN program on Theoretical Astroparticle Physics and by the World Premier 
International Research Center Initiative (WPI Initiative, MEXT), Japan.
The authors would like to thank Kavli IPMU, University of Tokyo, 
where part of this study was done, for the kind hospitality. 

%
\appendix
\section*{Appendix}
%
%
%
\section{Modular forms of $A_4$  with higher weights}
\label{modular-forms} 
%
%
The lowest weight 2 triplet modular forms of $A_4$ are given as:
\begin{align}
{\bf Y^{(\rm 2)}_3}
=\begin{pmatrix}Y_1\\Y_2\\Y_3\end{pmatrix}
=\begin{pmatrix}
1+12q+36q^2+12q^3+\dots \\
-6q^{1/3}(1+7q+8q^2+\dots) \\
-18q^{2/3}(1+2q+5q^2+\dots)\end{pmatrix}\,,
\label{Y(2)}
\end{align}
%
where $q=\exp[2\pi i \tau]$.
They satisfy also the constraint \cite{Feruglio:2017spp}:
\begin{align}
Y_2^2+2Y_1 Y_3=0~.
\label{condition}
\end{align}
%
For weight 4, five modular forms are given as:
\begin{align}
&\begin{aligned}
{\bf Y^{(\rm 4)}_1}=Y_1^2+2 Y_2 Y_3=E_4 \ , \qquad
{\bf Y^{(\rm 4)}_{1'}}=Y_3^2+2 Y_1 Y_2 \ , \qquad
{\bf Y^{(\rm 4)}_{1''}}=0 \, ,
\end{aligned}\nonumber \\
\nonumber \\
&\begin{aligned} {\bf Y^{(\rm 4)}_{3}}=
\begin{pmatrix}
Y_1^{(4)}  \\
Y_2^{(4)} \\
Y_3^{(4)}
\end{pmatrix}
=
\begin{pmatrix}
Y_1^2-Y_2 Y_3  \\
Y_3^2 -Y_1 Y_2 \\
Y_2^2-Y_1 Y_3
\end{pmatrix}\ , 
\end{aligned}
\label{weight4}
\end{align}
%
where ${\bf Y^{(\rm 4)}_{1''}}$ vanishes due to the 
constraint of Eq.\,(\ref{condition}).

For weigh 6, there are  seven modular forms as:
\begin{align}
&\begin{aligned}
{\bf Y^{(\rm 6)}_1}=Y_1^3+ Y_2^3+Y_3^3 -3Y_1 Y_2 Y_3=E_6  \ ,
\end{aligned} \nonumber \\
\nonumber \\
&\begin{aligned} {\bf Y^{(\rm 6)}_3}\equiv 
\begin{pmatrix}
Y_1^{(6)}  \\
Y_2^{(6)} \\
Y_3^{(6)}
\end{pmatrix}
=(Y_1^2+2 Y_2 Y_3)
\begin{pmatrix}
Y_1  \\
Y_2 \\
Y_3
\end{pmatrix}\ , \qquad
\end{aligned}
\begin{aligned} {\bf Y^{(\rm 6)}_{3'}}\equiv
\begin{pmatrix}
Y_1^{'(6)}  \\
Y_2^{'(6)} \\
Y_3^{'(6)}
\end{pmatrix}
=(Y_3^2+2 Y_1 Y_2 )
\begin{pmatrix}
Y_3  \\
Y_1 \\
Y_2
\end{pmatrix}\ . 
\end{aligned}
\label{weight6}
\end{align}
%
For weigh 8, there are  nine modular forms as:
\begin{align}
&\begin{aligned}
{\bf Y^{(\rm 8)}_1}=(Y_1^2+2Y_2 Y_3)^2=E_8=E_4^2 \, , \quad
{\bf Y^{(\rm 8)}_{1'}}=(Y_1^2+2Y_2 Y_3)(Y_3^2+2Y_1 Y_2)\, , \quad 
{\bf Y^{(\rm 8)}_{1"}}=(Y_3^2+2Y_1 Y_2)^2\,,
\end{aligned} \nonumber \\
\nonumber \\
&\begin{aligned} {\bf Y^{(\rm 8)}_3}\equiv \hskip -1 mm
\begin{pmatrix}
Y_1^{(8)}  \\
Y_2^{(8)} \\
Y_3^{(8)}
\end{pmatrix} \hskip -1 mm
=(Y_1^2+2 Y_2 Y_3) \hskip -1 mm
\begin{pmatrix}
Y_1^2-Y_2 Y_3  \\
Y_3^2 -Y_1 Y_2 \\
Y_2^2-Y_1 Y_3
\end{pmatrix} , \ \ \
\end{aligned}
\begin{aligned} {\bf Y^{(\rm 8)}_{3'}}\equiv \hskip -1 mm
\begin{pmatrix}
Y_1^{'(8)}  \\
Y_2^{'(8)} \\
Y_3^{'(8)}
\end{pmatrix} \hskip -1 mm
=(Y_3^2+2 Y_1 Y_2 ) \hskip -1 mm
\begin{pmatrix}
Y_2^2-Y_1 Y_3 \\
Y_1^2-Y_2 Y_3  \\
Y_3^2 -Y_1 Y_2 
\end{pmatrix}.
\end{aligned}
\label{weight8}
\end{align}
%
For weigh 10, there are  eleven modular forms as:
\begin{align}
&{\bf Y^{(\rm 10)}_1}=(Y_1^2+2Y_2 Y_3) (Y_1^3+ Y_2^3+Y_3^3 -3Y_1 Y_2 Y_3)=E_{10}=E_4 E_6 \, ,\nonumber\\
&{\bf Y^{(\rm 10)}_{1'}}=(Y_3^2+2Y_1 Y_2) (Y_1^3+ Y_2^3+Y_3^3 -3Y_1 Y_2 Y_3)\ ,
\nonumber \\
& {\bf Y^{(\rm 10)}_{3,{\rm 1}}}\equiv 
\begin{pmatrix}
Y_{1,1}^{(10)}  \\
Y_{2,1}^{(10)} \\
Y_{3,1}^{(10)}
\end{pmatrix} 
=(Y_1^2+2 Y_2 Y_3)^2 
\begin{pmatrix}
Y_1  \\
Y_2 \\
Y_3
\end{pmatrix} \,,\nonumber\\
& {\bf Y^{(\rm 10)}_{3,{\rm 2}}}\equiv 
\begin{pmatrix}
Y_{1,2}^{(10)}  \\
Y_{2,2}^{(10)} \\
Y_{3,2}^{(10)}
\end{pmatrix} 
=(Y_3^2+2 Y_1 Y_2 )^2 
\begin{pmatrix}
Y_2 \\
Y_3  \\
Y_1 
\end{pmatrix}\,,\nonumber\\
& {\bf Y^{(\rm 10)}_{3,{\rm 3}}}\equiv 
\begin{pmatrix}
Y_{1,3}^{(10)}  \\
Y_{2,3}^{(10)} \\
Y_{3,3}^{(10)}
\end{pmatrix} 
=(Y_1^2+2 Y_2 Y_3 )(Y_3^2+2 Y_1 Y_2 )
\begin{pmatrix}
Y_3 \\
Y_1  \\
Y_2 
\end{pmatrix}\,.
\label{weight10}
\end{align}
%
For weigh 12, there are  thirteen modular forms as:
\begin{align}
&\begin{aligned}
&{\bf Y^{(\rm 12)}_{1{\rm A}}}=(Y_1^2+2Y_2 Y_3)^3=E_4^3 \, , \ \  \qquad\qquad \
{\bf Y^{(\rm 12)}_{1{\rm B}}}=(Y_3^2+2Y_1 Y_2)^3=E_6^2-E_4^3 \, ,\nonumber\\
&{\bf Y^{(\rm 12)}_{1'}}=(Y_1^2+2Y_2 Y_3)^2(Y_3^2+2Y_1 Y_2)\, , \qquad 
{\bf Y^{(\rm 12)}_{1"}}=(Y_1^2+2Y_2 Y_3)(Y_3^2+2Y_1 Y_2)^2\,,
\end{aligned} \nonumber \\
\nonumber \\
&\begin{aligned} {\bf Y^{(\rm 12)}_3}\equiv \hskip -1 mm
\begin{pmatrix}
Y_1^{(12)}  \\
Y_2^{(12)} \\
Y_3^{(12)}
\end{pmatrix} \hskip -1 mm
=2(Y_1^2+2 Y_2 Y_3)^2 
\begin{pmatrix}
Y_1^2-Y_2 Y_3  \\
Y_3^2 -Y_1 Y_2 \\
Y_2^2-Y_1 Y_3
\end{pmatrix} , 
\end{aligned} \nonumber\\
&\begin{aligned} {\bf Y^{(\rm 12)}_{3'}}\equiv \hskip -1 mm
\begin{pmatrix}
Y_1^{'(12)}  \\
Y_2^{'(12)} \\
Y_3^{'(12)}
\end{pmatrix} \hskip -1 mm
=-2(Y_3^2+2 Y_1 Y_2 )^2 
\begin{pmatrix}
Y_3^2-Y_1 Y_2\\
Y_2^2-Y_1 Y_3  \\
Y_1^2 -Y_2 Y_3 
\end{pmatrix},
\end{aligned} \nonumber\\
&\begin{aligned} {\bf Y^{(\rm 12)}_{3'}}\equiv 
\begin{pmatrix}
Y_1^{''(12)}  \\
Y_2^{''(12)} \\
Y_3^{''(12)}
\end{pmatrix} 
=(Y_1^2+2 Y_2 Y_3)(Y_3^2+2 Y_1 Y_2 )
\begin{pmatrix}
Y_2^2-Y_1 Y_3 \\
Y_1^2-Y_2 Y_3  \\
Y_3^2 -Y_1 Y_2 
\end{pmatrix}.
\end{aligned}
\label{weight14}
\end{align}
For weigh 14, there are fifteen   modular forms as:
\begin{align}
&{\bf Y^{(\rm 14)}_1}=(Y_1^2+2Y_2 Y_3)^2
(Y_1^3+ Y_2^3+Y_3^3 -3Y_1 Y_2 Y_3)=E_4^2 E_6 \, , \nonumber\\
&{\bf Y^{(\rm 14)}_{1'}}=(Y_1^2+2Y_2 Y_3)(Y_3^2+2Y_1 Y_2) (Y_1^3+ Y_2^3+Y_3^3 -3Y_1 Y_2 Y_3)\, ,\nonumber\\
&{\bf Y^{(\rm 14)}_{1"}}=(Y_3^2+2Y_1 Y_2)^2 (Y_1^3+ Y_2^3+Y_3^3 -3Y_1 Y_2 Y_3)\,.
\end{align}
Four triplets are obtained  by 
${\bf Y^{(\rm 10)}_{3}}\otimes {\bf Y^{(\rm 4)}_{1}}$
and ${\bf Y^{(\rm 8)}_{3}}\otimes {\bf Y^{(\rm 6)}_{1}}$.

For weigh 16, there are  seventeen modular forms as:
\begin{align}
&{\bf Y^{(\rm 16)}_{1{\rm A}}}=(Y_1^2+2Y_2 Y_3)^4=E_4^4 \, , \qquad\qquad
{\bf Y^{(\rm 16)}_{1{\rm B}}}=(Y_1^2+2Y_2 Y_3)(Y_3^2+2Y_1 Y_2)^3
=E_4(E_6^2-E_4^3) \, ,\nonumber\\
&{\bf Y^{(\rm 16)}_{1'{\rm A}}}=(Y_1^2+2Y_2 Y_3)^3(Y_3^2+2Y_1 Y_2) \, , \quad
{\bf Y^{(\rm 16)}_{1'{\rm B}}}=(Y_3^2+2Y_1 Y_2)^4\, ,\nonumber\\
&{\bf Y^{(\rm 16)}_{1"}}=(Y_1^2+2Y_2 Y_3)^2(Y_3^2+2Y_1 Y_2)^2\, .  
\label{weight16}
\end{align}
Four triplets are obtained  by 
${\bf Y^{(\rm 12)}_{3}}\otimes {\bf Y^{(\rm 4)}_{1}}$
and ${\bf Y^{(\rm 10)}_{3}}\otimes {\bf Y^{(\rm 6)}_{1}}$.

The modular form $E_k$ is  the holomorphic normalized Eisenstein series with weight $k$,
which is given
\begin{align}
E_k(\tau)=\frac{1}{{2 \zeta(k)}} \sum_{(m,n)\not =(0.0)}
\frac{1}{(m+n\tau)^k} \,,
\label{Eisen}
\end{align}
where $m$ and $n$ are integers.


We show the values of singlets of modular form
at the fixed points, $i$, $\omega$, $i\infty$ in
Table \ref{tb:singlets-fixed-points}.
\begin{table}[H]
	\centering
	\begin{tabular}{|c|c|c|c|c|} \hline 
		\rule[14pt]{0pt}{1pt}
		$k$ & $\bf r$	& $\tau_0=i$ &$\tau_0=\omega$ &	$\tau_0=i\infty$\\ \hline 
		\rule[14pt]{0pt}{2pt}
		$2$& ${\bf 3}\,(Y_1,Y_2,Y_3)$	& $Y_0\,(1,1-\sqrt{3}, -2+\sqrt{3})$ &$Y_0\,(1,\omega, -\frac{1}{2}\omega^2)$
		& $Y_0^2\,(1,0, 0)$\\ \hline
		\rule[14pt]{0pt}{2pt}
		$4$& $\{\bf 1,\ 1'\}$	& $Y_0^2\,\{(6\sqrt{3}-9), -(6\sqrt{3}-9) \}$ &$Y_0^2\,(0,\ \frac94\omega)$
		& $Y_0^2\,\{1,\ 0\}$\\ \hline
		\rule[14pt]{0pt}{2pt}
		$6$	&$\bf 1$	&  0 &$ Y_0^3\,\frac{27}{8}$
		& $ Y_0^3$\\ \hline
		\rule[14pt]{0pt}{2pt}
		$8$	&$\bf 1$	&   $Y_0^4\,27\,(7-4\sqrt{3})$ &0
		& $ Y_0^4$\\ 
		\rule[14pt]{0pt}{2pt}
		&$\bf 1'$	&   $-Y_0^4\,27\,(7-4\sqrt{3})$ &0
		& 0\\ 
		\rule[14pt]{0pt}{2pt}
		&$\bf 1''$& $Y_0^4\,27\,(7-4\sqrt{3})$ &$ Y_0^4\,\frac{81}{16}\omega^2$
		& 0\\ \hline
		\rule[14pt]{0pt}{2pt}
		$10$& $\{\bf 1,\ 1'\}$	& $Y_0^5\,\{0,\ 0 \}$ 
		&$Y_0^5\,( 0,\ \frac{243}{32}\omega)$
		& $Y_0^5\,\{1,\ 0\}$\\ \hline
		\rule[14pt]{0pt}{2pt}
		$12$& $\{\bf 1_{\rm A},\ 1_{\rm B}\}$
		& $81Y_0^6\,\{\,(26\sqrt{3}-45), -(26\sqrt{3}-45) \}$ 
		&$Y_0^6\,\{ 0,\ \frac{729}{64}\}$
		& $Y_0^6\,\{1,\ 0\}$\\
		\rule[14pt]{0pt}{2pt}
		& $\{\bf 1',\ 1''\}$
		& $-81Y_0^6\,\{(26\sqrt{3}-45),  -(26\sqrt{3}-45)\}$ 
		&\{0,\ 0\}	&\{0,\ 0\}\\ \hline
		\rule[14pt]{0pt}{2pt}
		$14$& $\{\bf 1,\ 1',\ 1"\}$
		& $\{0,\ 0,\ 0\}$ 
		&$Y_0^7\,\{0,\ 0,\ \frac{2187}{128} \omega^2\}$
		& $Y_0^7\,\{1,\ 0,\ 0\}$\\\hline
		\rule[14pt]{0pt}{2pt}
		$16$& $\{\bf 1_{\rm A},\ 1_{\rm B}\}$
		&  $729Y_0^8\,\{-(56\sqrt{3}-97),  (56\sqrt{3}-97)\}$ 
		&$\{ 0,\ 0\}$
		& $Y_0^8\,(1,\ 0)$\\
		\rule[14pt]{0pt}{2pt}
		& $\{\bf 1'_A,\ 1'_B\}$
		& $729Y_0^8\,\{(56\sqrt{3}-97),  -(56\sqrt{3}-97)\}$ 
		&$Y_0^8\,\{0,\  \frac{6561}{256} \omega\}$	&\{0,\ 0\}\\ 
		\rule[14pt]{0pt}{2pt}
		& $1"$
		&  $-729Y_0^8\,(56\sqrt{3}-97)$ 
		&0	&0\\ \hline
		\rule[14pt]{0pt}{2pt}
		& $Y_0=Y_1(\tau_0)$	& $Y_1(i)=1.0225...$ & $Y_1(\omega)=0.948...$
		&$Y_1(i\infty)=1$\\ \hline			
	\end{tabular}
	\caption{Modular forms of singlets with  weight  $k=4,\,6,\,8,\,10,\,12,\,14,\,16$,  at  the fixed point $\tau_0$.
	}
	\label{tb:singlets-fixed-points}
\end{table}

%
\section{Input data of quark masses and CKM elements}
\label{Appen-inputs} 
%

The  modulus $\tau$ breaks the modular invariance
by obtaining a VEV at some high mass scale.
We assume this to be the GUT scale. Correspondingly, the values of the
quark masses and CKM parameters at the GUT scale play the role of
the observables that have to be reproduced by the considered quark flavour
models. They are obtained using the renormalisation group
(RG) equations which describe the ``running'' of the observables of interest
from the electroweak scale, where they are measured, to the GUT scale.
In the analyses which follow we adopt the numerical values
of the quark Yukawa couplings 
at the GUT scale $2\times 10^{16}$ GeV
derived in the framework of the minimal SUSY breaking scenarios
with $\tan\beta=5$ 
\cite{Antusch:2013jca}:
\begin{align}
\begin{aligned}
&\frac{y_d}{y_b}=9.21\times 10^{-4}\ (1\pm 0.111) \, , 
\qquad
\frac{y_s}{y_b}=1.82\times 10^{-2}\ (1\pm 0.055) \, , \\
\rule[15pt]{0pt}{1pt}
&\frac{y_u}{y_t}=5.39\times 10^{-6}\ (1\pm 0.311)\,  , 
\qquad
\frac{y_c}{y_t}=2.80\times 10^{-3}\ (1\pm 0.043)\,. \\
\end{aligned}
\label{Datamass}
\end{align}
%
The quark masses are given as $m_q=y_q v_H$ with $v_H=174$ GeV.
The choice of relatively small value of $\tan\beta$ allows us to 
avoid relatively large $\tan\beta$-enhanced threshold corrections in 
the RG running of the Yukawa couplings. 
We set these corrections to zero.

{ The quark flavour mixing is given by the CKM matrix,
	which has  three independent mixing angles and one CP violating phase.
	These mixing angles are given by 
	the absolute values of the three CKM elements
$|V_{us}^{\rm }|$, $|V_{cb}^{\rm }|$ and $|V_{ub}^{\rm }|$.
We take the present data on the three CKM elements
in Particle Data Group (PDG) edition 
of Review of Particle Physics 
\cite{ParticleDataGroup:2022pth} as:
}
\begin{align}
\begin{aligned}
|V_{us}^{\rm }|=0.22500\pm 0.00067 \, , \quad
|V_{cb}^{\rm }|=0.04182^{\pm 0.00085}_{-0.00074} \,,  \quad
|V_{ub}^{\rm }|=0.00369\pm 0.00011\, .
\end{aligned}\label{DataCKM}
\end{align}
%
By using these values as input and $\tan\beta=5$ we obtain 
the CKM 
mixing angles  at the GUT scale of $2\times 10^{16}$ GeV
\cite{Antusch:2013jca}:
\begin{align}
|V_{us}^{\rm }|=0.2250\,(1\pm 0.0032) \, , \quad
|V_{cb}^{\rm }|=0.0400\, (1\pm 0.020) \,,  \quad
|V_{ub}^{\rm }|=0.00353\,(1\pm 0.036)\,.
\label{DataCKM-GUT}
\end{align}

The tree-level decays of $B\to D^{(*)}K^{(*)}$ are used as the standard candles
of the CP violation. 
{ The latest world average of the CP violating phase 
is given in PDG2022 \cite{ParticleDataGroup:2022pth} as:
}
\begin{equation}
\delta_{CP}={66.2^\circ}^{+ 3.4^\circ}_{-3.6^\circ}\,. 
\label{CKMphase}
\end{equation}
%
Since the phase is almost independent of the evolution of RG equations,
we refer to this value in the numerical discussions.
The rephasing invariant CP violating measure $J_{\rm CP}$ \cite{Jarlskog:1985ht}
is also given in \cite{ParticleDataGroup:2022pth}:
\begin{equation}
J_{\rm CP}=3.08^{+0.15}_{-0.13} \times 10^{-5} \,.
\label{JCP}
\end{equation}
%
Taking into account the RG effects on the mixing angles 
for $\tan\beta = 5$, we have at the GUT scale $2\times 10^{16}$ GeV: 
\begin{equation}
J_{\rm CP}= 2.80^{+0.14}_{-0.12}\times 10^{-5}\,.
\label{JCPGUT}
\end{equation}

%
\section{Quark mass matrices with texture zeros}
\label{Appen-texture}
%

In this Appendix, we present three numerical examples
of the quark mass matrix with  the texture zeros,
which are consistent with the observed masses and CKM elements
given  in Appendix \ref{Appen-inputs}.
 
%
\subsection{Texture (1)}
\label{Appen-texture-1}
%
%

{ Consider the quark mass matrices with the texture zeros of model (1):} 
\begin{align}
&  M_Q =v_Q
\begin{pmatrix}
0  & 0 & a_Q  \\
0&b_Q & c_Q \\
d_Q& e_Q& f_Q  \, e^{i \varphi_Q}\end{pmatrix}_{RL},
 \qquad Q=D, U\ ,
\label{texture-1}
\end{align}
%
where $a_D,\, b_D,\, c_D,\, d_D,\, e_D,\,f_D,\,a_U,\, b_U,\, c_U,\, d_U,\, e_U,\,f_U$ are real constants, and 
$ \phi_D,\, \phi_U$ are CP violating (CPV) phases.
The determinants of the 
mass matrices $M_D$ and $M_U$ are real and are given in terms of 
the real parameters  $a_Q$, $b_Q$ and $d_D$:
\begin{align}
{\rm det }\,[M_D]=-a_D b_D d_D\,,\qquad\qquad 
{\rm det }\, [M_U]=-a_U b_U d_U\,.
\label{det1}
\end{align}
%
If these mass matrices keep strictly their zero structures  
and describe correctly the observed CP violation, 
they are candidates for solving the strong CP problem.

{ Since the number of parameters in $M_D$ and $M_U$ are 14,  
four parameters are redundant from the point of view of 
reproducing the six quark masses and the four independent 
CKM mixing angles and CPV phase (10 observables ).
We present the mass matrices in the case of 
$|a_Q|\sim |b_Q|  \sim |d_Q|$.
Using as input the  observed quark masses and the 
CKM parameters and performing a statistical analysis
we show a typical numerical example of the mass matrices 
which reproduce the quark data in Appendix \ref{Appen-inputs}:}
\begin{align}
&  M_D \sim 
\begin{pmatrix}
0  & 0 & 1  \\
0&0.901 & -11.63 \\
0.105& -0.695& 12.96 \, e^{i 24.36^\circ}\end{pmatrix},
\ \
M_U \sim
\begin{pmatrix}
0  & 0 & 1  \\
0&1.23 & -80.80 \\
-0.586& 9.63& 386.81  \, e^{-i 150.26} \\
\end{pmatrix},
\label{texture-N1}
\end{align}
%
{where $|a_Q| \sim |b_Q|  \sim |d_Q|$ are approximately 
satisfied  with only  $|d_D|$ being somewhat smaller 
than the other five constants 
\footnote{The magnitudes of the constants, including   
$c_Q$, $e_Q$ and $f_Q$, $Q=D,U$, which in the 
modular $A_4$ model  multiply the modular forms present in 
$M_D$ and $M_U$, are discussed in section \ref{sec:fits}.
}. 
The matrices $M_D$ and $M_U$ 
with the numerical values of the real constants and 
of the CPV phases $\varphi_D=24.13^\circ$ and $\varphi_U=-150.26^\circ$
as given in Eq.\,(\ref{texture-N1}), 
provide a good quality of the fit of the 
quark mass ratios, the CKM mixing angles and 
the CPV phase $\delta_{\rm CP}$
($N\sigma=0.840$ C.L.), as seen in Table \ref{tab:texture1}.
 }
\begin{table}[H]
	\small{
		\begin{center}
			\renewcommand{\arraystretch}{1.1}
			\begin{tabular}{|c|c|c|c|c|c|c|c|c|c|} \hline
				\rule[14pt]{0pt}{3pt}  
				& $\frac{m_s}{m_b}\hskip -1 mm\times\hskip -1 mm 10^2$ 
				& $\frac{m_d}{m_b}\hskip -1 mm\times\hskip -1 mm 10^4$& $\frac{m_c}{m_t}\hskip -1 mm\times\hskip -1 mm 10^3$&$\frac{m_u}{m_t}\hskip -1 mm\times\hskip -1 mm 10^6$&
				$|V_{us}|$ &$|V_{cb}|$ &$|V_{ub}|$&$|J_{\rm CP}|$& $\delta_{\rm CP}$
				\\
				\hline
				\rule[14pt]{0pt}{3pt}  
				Fit &$1.86$ & $ 9.52$
				& $2.80$ & $4.18$&
				$0.2249$ & $0.0400$ & $0.00355$ &
				$2.86\hskip -1 mm\times\hskip -1 mm 10^{-5}$&$67.0^\circ$
				\\ \hline
				\rule[14pt]{0pt}{3pt}
				Exp	 &$1.82$ & $9.21$ 
				& $2.80$& $ 5.39$ &
				$0.2250$ & $0.0400$ & $0.00353$ &$2.80\hskip -1 mm\times\hskip -1 mm 10^{-5}$&$66.2^\circ$\\
				$1\,\sigma$	&$\pm 0.10$ &$\pm 1.02$ & $\pm 0.12$& $\pm 1.68$ &$ \pm 0.0007$ &
				$ \pm 0.0008$ & $ \pm 0.00013$ &$^{+0.14}_{-0.12}\hskip -1 mm\times \hskip -1 mm 10^{-5}$&
				$^{+ 3.4^\circ}_{-3.6^\circ}$\\ \hline  
			\end{tabular}
		\end{center}
		\caption{Results of the fits of the quark mass ratios, 
			CKM mixing angles, $J_{\rm CP}$ and $\delta_{\rm CP}$. 'Exp' denotes the  values of the observables 
			at the GUT scale, including $1\sigma$ error.
		}
		\label{tab:texture1}
	}
\end{table}
%

%
\subsection{Texture (2)}
\label{Appen-texture-2}
%
%
{ 
Consider next the quark mass matrices with the texture zeros of model (2):} 
\begin{align}
&  M_Q =v_Q
\begin{pmatrix}
0  & 0 & a_Q  \\
b_Q& 0 & c_Q\\
e_Q& d_Q& f_Q  \, e^{i \varphi_Q}\end{pmatrix}_{RL},
\qquad Q=D, U\ ,
\label{texture-2}
\end{align}
%
where again the constants 
$a_D,\, b_D,\, c_D,\, d_D,\, e_D,\,f_D,\,a_U,\, b_U,\, c_U,\, d_U,\, e_U,\,f_U$ 
are real, and $ \varphi_D,\, \varphi_U$ are CPV phases.
 The determinants of 
the mass matrices $M_D$ and $M_U$ are 
given in terms of $a_Q,\, b_Q,\,  d_D$
and  real:
\begin{align}
{\rm det }\,[M_D]=a_D b_D d_D\,,\qquad\qquad 
{\rm det }\, [M_U]=a_U b_U d_U\,.
\label{det2}
\end{align}
%
If these mass matrices keep their 
zero elements  strictly zero and describe the 
the observed CP violation, 
they are candidates for solving the strong CP problem.

 Performing a statistical analysis 
we have found  a typical numerical example of the mass matrices
$M_D$ and $M_U$ which describe the quark data
 in Appendix \ref{Appen-inputs}: 
\begin{align}
&  M_D \sim 
\begin{pmatrix}
0  & 0 & 1  \\
0.0610&0 & 4.576 \\
0.0380&0.733& 12.901  \, e^{i\,104.836^\circ}\end{pmatrix},
\ \
M_U \sim
\begin{pmatrix}
0  & 0 & 1  \\
0.005752&0 & 12.44 \\
0.00290&1.625& 83.89  \, e^{i\, 56.64} \\
\end{pmatrix}.
\label{texture-N2}
\end{align}
%
{ The quark mass ratios and the CKM matrix obtained by diagonalising 
the matrices $M_D$ and $M_U$ given in Eq.\,(\ref{texture-N2})  
are completely consistent with observed one including
the CP violating phase  $\delta_{\rm CP}$ ($N\sigma=0.217$ C.L.),
as seen in Table \ref{tab:texture2}.
}
%
\begin{table}[H]
	\small{
		\begin{center}
			\renewcommand{\arraystretch}{1.1}
			\begin{tabular}{|c|c|c|c|c|c|c|c|c|c|} \hline
				\rule[14pt]{0pt}{3pt}  
				& $\frac{m_s}{m_b}\hskip -1 mm\times\hskip -1 mm 10^2$ 
				& $\frac{m_d}{m_b}\hskip -1 mm\times\hskip -1 mm 10^4$& $\frac{m_c}{m_t}\hskip -1 mm\times\hskip -1 mm 10^3$&$\frac{m_u}{m_t}\hskip -1 mm\times\hskip -1 mm 10^6$&
				$|V_{us}|$ &$|V_{cb}|$ &$|V_{ub}|$&$|J_{\rm CP}|$& $\delta_{\rm CP}$
				\\
				\hline
				\rule[14pt]{0pt}{3pt}  
				Fit &$1.87$ & $ 9.21$
				& $2.82$ & $5.43$&
				$0.2249$ & $0.0400$ & $0.00352$ &
				$2.82\hskip -1 mm\times\hskip -1 mm 10^{-5}$&$66.3^\circ$
				\\ \hline
				\rule[14pt]{0pt}{3pt}
				Exp	 &$1.82$ & $9.21$ 
				& $2.80$& $ 5.39$ &
				$0.2250$ & $0.0400$ & $0.00353$ &$2.80\hskip -1 mm\times\hskip -1 mm 10^{-5}$&$66.2^\circ$\\
				$1\,\sigma$	&$\pm 0.10$ &$\pm 1.02$ & $\pm 0.12$& $\pm 1.68$ &$ \pm 0.0007$ &
				$ \pm 0.0008$ & $ \pm 0.00013$ &$^{+0.14}_{-0.12}\hskip -1 mm\times \hskip -1 mm 10^{-5}$&
				$^{+ 3.4^\circ}_{-3.6^\circ}$\\ \hline  
			\end{tabular}
		\end{center}
		\caption{Results of the fits of the quark mass ratios, 
			CKM mixing angles, $J_{\rm CP}$ and $\delta_{\rm CP}$. 'Exp' denotes the  values of the observables 
			at the GUT scale, including $1\sigma$ error.
		}
		\label{tab:texture2}
	}
\end{table}

%
\subsection{Texture (3)}
\label{Appen-texture-3}
%
%
{ Let us analyse finally  the quark mass matrices with the texture zeros 
of model (3):} 
\begin{align}
&  M_Q =v_Q
\begin{pmatrix}
a_Q &0 & 0  \\
c_Q&  b_Q &0\\
f_Q \, e^{i \varphi_Q}& e_Q& d_Q \end{pmatrix}_{RL},
\qquad Q=D, U\ ,
\label{texture-3}
\end{align}
%
where as in the previous two cases the constants
 $a_D,\, b_D$, $c_D,\, d_D$, $e_D,\,f_D$, $a_U,\, b_U$, 
$c_U,\, d_U$, $e_U,\,f_U$ 
are real, and $ \varphi_D,\, \varphi_U$ are CPV phases.
{ Also in this case the determinants of the quark mass matrices 
$M_D$ and $M_U$ are given in terms of real parameters, namely of  
$a_Q,\,b_Q,\,d_D$, and are real:
}
\begin{align}
{\rm det }\,[M_D]=a_D b_D d_D\,,\qquad\qquad 
{\rm det }\, [M_U]=a_U b_U d_U\,,
\label{det3}
\end{align}
%
As in the previous two cases we note that 
if the zero elements of these mass matrices  
remain strictly zero, and the mass matrices describe 
correctly the observed quark masses, and especially 
the quark mixing angles and CP violation in the quark sector, 
they are a candidate for solving the strong CP problem.

{We will present next a typical numerical example of the mass matrices 
which describe the quark data.
}
Since the number of parameters is 14,  four parameters are
redundant for reproducing the quark masses 
and CKM elements (10 observables).
{In this case, following  Ref.\cite{Tanimoto:2016rqy}, 
we set $c_U=e_Q=f_U=0$ (also $\varphi_U=0$). 
Performing a statistical analysis we get  
a good description of the quark data in Appendix \ref{Appen-inputs} with:
}
\begin{align}
&  M_D \sim 
\begin{pmatrix}
0.000948  & 0 & 0  \\
0.004209 &0.01828 &0 \\
0.003519  \, e^{i\,66.25^\circ}&0.0400& 1 \end{pmatrix},
\ \
M_U \sim
\begin{pmatrix}
5.39 \times 10^{-6}  & 0 & 0  \\
0&2.80 \times  10^{-3} & 0 \\
0&0& 1   \\
\end{pmatrix}.
\label{texture-N3}
\end{align}

%
 The quark mass ratios, and especially the 
CKM mixing angles, the $J_{\rm CP}$ factor and the CPV phase $\delta_{\rm CP}$ 
obtained using the numerical matrices given in Eq.\,(\ref{texture-N3})
are completely consistent with observed one 
($ N\sigma=0.113$ C.L.), as seen in Table \ref{tab:texture3}.
\begin{table}[H]
	\small{
		\begin{center}
			\renewcommand{\arraystretch}{1.1}
			\begin{tabular}{|c|c|c|c|c|c|c|c|c|c|} \hline
				\rule[14pt]{0pt}{3pt}  
				& $\frac{m_s}{m_b}\hskip -1 mm\times\hskip -1 mm 10^2$ 
				& $\frac{m_d}{m_b}\hskip -1 mm\times\hskip -1 mm 10^4$& $\frac{m_c}{m_t}\hskip -1 mm\times\hskip -1 mm 10^3$&$\frac{m_u}{m_t}\hskip -1 mm\times\hskip -1 mm 10^6$&
				$|V_{us}|$ &$|V_{cb}|$ &$|V_{ub}|$&$|J_{\rm CP}|$& $\delta_{\rm CP}$
				\\
				\hline
				\rule[14pt]{0pt}{3pt}  
				Fit &$1.87$ & $ 9.23$
				& $2.80$ & $5.39$&
				$0.2250$ & $0.0400$ & $0.00352$ &
				$2.82\hskip -1 mm\times\hskip -1 mm 10^{-5}$&$66.2^\circ$
				\\ \hline
				\rule[14pt]{0pt}{3pt}
				Exp	 &$1.82$ & $9.21$ 
				& $2.80$& $ 5.39$ &
				$0.2250$ & $0.0400$ & $0.00353$ &$2.80\hskip -1 mm\times\hskip -1 mm 10^{-5}$&$66.2^\circ$\\
				$1\,\sigma$	&$\pm 0.10$ &$\pm 1.02$ & $\pm 0.12$& $\pm 1.68$ &$ \pm 0.0007$ &
				$ \pm 0.0008$ & $ \pm 0.00013$ &$^{+0.14}_{-0.12}\hskip -1 mm\times \hskip -1 mm 10^{-5}$&
				$^{+ 3.4^\circ}_{-3.6^\circ}$\\ \hline  
			\end{tabular}
		\end{center}
		\caption{Results of the fits of the quark mass ratios, 
			CKM mixing angles, $J_{\rm CP}$ and $\delta_{\rm CP}$. 
'Exp' denotes the  values of the observables 
at the GUT scale, including $1\sigma$ error.
		}
		\label{tab:texture3}
	}
\end{table}


\end{document}